\documentclass{scrartcl}

\usepackage{stmaryrd}
\usepackage{xspace}
\usepackage{color}
\usepackage{natbib}
\usepackage{tikz}
\usetikzlibrary{matrix,arrows,shapes,calc,snakes}
\usepackage{amsthm}

\theoremstyle{definition}
\newtheorem{example}{Example}

\usepackage{epigram}

\usepackage{epigram-desc}
\usepackage{epigram-enum}
\usepackage{epigram-ornament}

\usepackage{epigram-nat}
\usepackage{epigram-list}
\usepackage{epigram-vec}
\usepackage{epigram-fin}
\usepackage{epigram-id}
\usepackage{epigram-bool}
\usepackage{epigram-maybe}
\usepackage{epigram-inverse}
\usepackage{epigram-function}

\newlength{\rulevgap}
\newlength{\ruleheight}
\newlength{\ruledepth}
\newsavebox{\rulebox}
\newlength{\GapLength}

\ColourEpigram

\newcommand{\trashtalk}[1]{\red{#1}}
\renewcommand{\trashtalk}[1]{}
\newcommand{\ie}{i.e.\ }
\newcommand{\eg}{e.g.\ }

\title{Transporting Functions across Ornaments \\
       \Large{Technical Report}}
\author{Pierre-Evariste Dagand
        \and Conor McBride}
\date{}
\begin{document}

\maketitle

\begin{abstract}
Programming with dependent types is a blessing and a curse. It is a
blessing to be able to bake invariants into the definition of
datatypes: we can finally write correct-by-construction
software. However, this extreme accuracy is also a curse: a datatype
is the combination of a structuring medium together with a special
purpose logic. These domain-specific logics hamper any effort of code
reuse among similarly structured data. In this paper, we exorcise our
datatypes by adapting the notion of ornament to our universe of
inductive families. We then show how code reuse can be achieved by
ornamenting functions. Using these functional ornament, we capture the
relationship between functions such as the addition of natural numbers
and the concatenation of lists. With this knowledge, we demonstrate
how the implementation of the former informs the implementation of the
latter: the user can ask the definition of addition to be lifted to
lists and she will only be asked the details necessary to carry on
adding lists rather than numbers. Our presentation is formalised in a
type theory with a universe of datatypes and all our constructions
have been implemented as generic programs, requiring no extension to
the type theory.
\end{abstract}


\section{Introduction}


Imagine designing a library for a ML-like language. For instance, we
start with natural numbers and their operations, then we move to
binary trees, then rose trees, etc. It is the garden of Eden:
datatypes are data-\emph{structures}, each coming with its optimised
set of operations. If, tempted by a snake, we move to a language with
richer datatypes, such as a dependently typed language, we enter the
Augean stables. Where we used to have binary trees, now we have
complete binary trees, red-black trees, AVL trees, and countless other
variants. Worse, we have to duplicate code across these
tree-like datatypes: because they are defined upon this common
binarily branching structure, a lot of computationally identical
operations will have to be duplicated for the type-checker to be
satisfied.

Since the ML days, datatypes have evolved: besides providing an
organising \emph{structure} for computation, they are now offering
more \emph{control} over what is a valid result. With richer datatypes,
the programmer can enforce invariants on top of the
data-structures. In such a system, programmers strive to express the
correctness of programs in their types: a well typed program is
correct \emph{by construction}, the proof of correctness being reduced
to type-checking.

A simple yet powerful recipe to obtain these richer datatypes is to
\emph{index} the data-structure. These datatypes have originally been
studied in the context of type theory under the name of
\emph{inductive families}~\citep{dybjer:inductive-families,
  morris:spf}. Inductive families made it to mainstream functional
programming with Generalised Algebraic Data-Types~\citep{cheney:gadt,
  xi:gadt}, a subset of inductive families for which type inference
remains decidable. Refinement
types~\citep{freeman:refinement,swamy:fstar} are another technique to
equip data-structures with rich
invariants. \citet{atkey:inductive-refinement} have shown how
refinement types relate to inductive families, and \citet{bernardy:realizability} establish
a connection with realisability.



However, these carefully crafted datatypes are a threat to any
library design: the same data-\emph{structure} is used for logically
incompatible purposes. This explosion of specialised datatypes is
overwhelming: these objects are too specialised to fit in a global
library. Yet, because they share this common structure, many
operations on them are extremely similar, if not exactly the same.
%
%
%
To address this issue, \citet{mcbride:ornament} developed
\emph{ornaments}, describing how one datatype can be enriched into
others \emph{with the same structure}. Such structure-preserving
transformations take two forms: one can \emph{extend} the initial type
with more information -- such as obtaining \(\Maybe{A}\) from
\(\Bool\) or \(\List{A}\) from \(\Nat\):
{
\[
\begin{array}{c@{\:\:}c@{\:\:}c}
\BoolDef & \stackrel{\OrnamentOf{\Maybe{}}}{\Longrightarrow} & \MaybeDef \\
\\
\NatDef & \stackrel{\OrnamentOf{\List{}}}{\Longrightarrow} & \ListDef
\end{array}
\]
}
Or one can \emph{refine} the indexing of the initial type by a finer
discipline -- e.g., obtaining \(\Fin\) by indexing \(\Nat\) with a
bound \(n\):
{
\[
\begin{array}{c@{\:\:\stackrel{\OrnamentOf{\Fin}}{\Longrightarrow}\:\:}c}
\NatDef & \FinDef
\end{array}
\]
}

One can also do both at the same time -- such as extending \(\Nat\)
into a \(\List{A}\) while refining the index to match the length of
the list:
{
\[
\begin{array}{c@{\:\:\stackrel{\OrnamentOf{\Vector{}}}{\Longrightarrow}\:\:}c}
\NatDef & \VectorDefEquality
\end{array}
\]}
Note that we declare datatype parameters \(\Param{\Var{A}}{\Set}\) in
brackets and datatype indices \(\Index{\Var{n}}{\Nat}\) in parentheses.
We make equational constraints on the latter only when needed,
and explicitly.



Because of their constructive nature, ornaments are not merely
identifying similar structures: they give an effective recipe to build
new datatypes from old, guaranteeing by construction that the structure is
preserved. Hence, we can obtain a plethora of new datatypes with
minimal effort.

\newcommand{\Plus}{\mathop{\green{+}}}
\newcommand{\Minus}{\mathop{\green{-}}}
\newcommand{\Le}{\mathop{\green{<}}}

\Spacedcommand{\Append}{\mathop{\green{+\!+}}}
\Spacedcommand{\Drop}{\Function{drop}}
\Spacedcommand{\Lookup}{\Function{lookup}}

Whilst we now have a good handle on the transformation of individual
datatypes, we are still facing a major reusability issue: a datatype often
comes equipped with a set of operations. Ornamenting this datatype, we
have to entirely re-implement many similar operations. For example,
the datatype \(\Nat\) comes with operations such as addition and
subtraction. When defining \(\List{A}\) as an ornament of \(\Nat\), it
seems natural to transport some structure-preserving function of
\(\Nat\) to \(\List{A}\), such as moving from addition of natural
numbers to concatenation of lists:
{
\[
\begin{array}{l@{}c@{\:\:}l}
\Code{
\Let{\PiTel{\Var{m}}{\Nat} & \Plus & \PiTel{\Var{n}}{\Nat}}{\Nat}{
\Return{\Zero & \Plus & \Var{n}}{\Var{n}}
\Return{(\Suc[\Var{m}]) & \Plus & \Var{n}}{\Suc[(\Var{m} \Plus \Var{n})]}
}} & 
\Rightarrow &
\Code{
\Let{\PiTel{\Var{xs}}{\List{\Var{A}}} &
         \Append &
         \PiTel{\Var{ys}}{\List{\Var{A}}}}
    {\List{\Var{A}}}{
\Return{\Nil & \Append & \Var{ys}}
       {\Var{ys}}
\Return{(\Cons[\Var{a}\: \Var{xs}]) & \Append & \Var{ys}}
       {\Cons[\Var{a}\: (\Var{xs} \Append \Var{ys})]}}
}
\\
\\
\Code{
\Let{\PiTel{\Var{m}}{\Nat} & \Minus & \PiTel{\Var{n}}{\Nat}}{\Nat}{
\Return{\Zero & \Minus & \Var{n}}{\Zero}
\Return{\Var{m} & \Minus & \Zero}{\Var{m}}
\Return{(\Suc[\Var{m}]) & \Minus & (\Suc[\Var{n}])}{\Var{m} \Minus \Var{n}}
}} & \Rightarrow &
\Code{
\Let{\Drop & 
         \PiTel{\Var{xs}}{\List{\Var{A}}} &
         \PiTel{\Var{n}}{\Nat}}
    {\List{\Var{A}}}{
\Return{\Drop & \Nil & \Var{n}}
       {\Nil}
\Return{\Drop & \Var{xs} & \Zero}
       {\Var{xs}}
\Return{\Drop & (\Cons[\Var{a}\: \Var{xs}]) & (\Suc[\Var{n}])}
       {\Drop[\Var{xs}\: \Var{n}]}
}}
\end{array}
\]}



More interestingly, the function we start with may involve several
datatypes, each of which may be ornamented differently. In this
paper, we develop the notion of \emph{functional ornament} as a
generalisation of ornaments to functions:
\begin{itemize}
\item We adapt ornaments to our universe of
  datatypes~\citep{dagand:levitation} in
  Section~\ref{sec:univ-data-orn}. This presentation benefits greatly
  from our ability to inspect indices when defining
  datatypes. This allows us to consider ornaments which
  \emph{delete} index-determined information, yielding
  a key simplification in the construction of an algebraic
  ornament from an ornamental algebra~;
\item We describe how functions can be transported through functional
  ornaments: `deletion' allows us a contrasting approach to
  \citet{ko:modularising-inductive}, internalising proof obligations.
  First, we manually work through an example in
  Section~\ref{sec:example-manual}. Then, we formalise the concept of
  functional ornament by a universe construction in
  Section~\ref{sec:lift}. Based on this universe, we establish the
  connection between a base function (such as \(\_\Plus\_\) and
  \(\_\Minus\_\)) and its ornamented version (such as, respectively,
  \(\_\Append\_\) and \(\Drop\)). Within this framework, we redevelop
  the example of Section~\ref{sec:example-manual} with all the
  automation offered by our constructions ;
\item In Section~\ref{sec:clever-constructors}, we provide further
  support to drive the computer into lifting functions
  semi-automatically. As we can see from our examples above, the lifted
  functions often follow the same recursion pattern and return similar
  constructors: with a few constructions, we shall remove further
  clutter and code duplication in our libraries.  
\item Finally, we add a generic gadget to our reusability kit in
  Section~\ref{sec:remove-index-computation}: we show how a careful
  use of an adjunction can absorb computation at the index-level --
  which are notoriously difficult to deal with -- and replace them by
  a special-purpose indexed family -- which is much easier to deal
  with.
\end{itemize}

This paper is an exercise in constructive mathematics: upon
identifying an isomorphism, we shall look at it with our
constructive glasses and obtain an effective procedure that lets us
cross the isomorphism.

We shall write our code in a syntax inspired by the
Epigram~\cite{mcbride:view-from-left} programming language. In
particular, we make use of the \emph{by} (\(\DoBy\)) and \emph{return}
(\(\DoReturn\)) programming gadgets, further extending them to account for
the automatic lifting of functions. For brevity, we write
pattern-matching definitions when the recursion pattern is evident and
unremarkable. Like ML, unbound variables in type
definitions are universally quantified, further abating
syntactic noise. The syntax of datatype definitions draws upon the ML
tradition as well: its novelty will be presented by way of examples in
Section~\ref{sec:univ-data-orn}. All the constructions presented in
this paper have been modelled in Agda, using only standard inductive
definitions and two levels of universes. The formalisation is available on Dagand's website.


\section{From \(\_ \Le \_\) to \(\Lookup\), manually}
\label{sec:example-manual}


There is an astonishing resemblance between the comparison function
\(\_ \Le \_\) on natural numbers and the list \(\Lookup\) function
(Fig.~\ref{fig:le-lookup}). 
Interestingly, the similarity is not merely at the level of types but
also in their implementation: their definitions follow the same
pattern of recursion (first, case analysis on the second element; then
induction on the first element) and they both return a failure value
(\(\False\) and \(\Nothing\) respectively) in the first case analysis
and a success value (\(\True\) and \(\Just\) respectively) in the base
case of the induction. 

\begin{figure}[t]
{
\[
\begin{array}{l@{\:\:\stackrel{?}{\Longrightarrow}\:\:}l}
\Let{\PiTel{\Var{m}}{\Nat} & \Le & \PiTel{\Var{n}}{\Nat}}{\Bool}{
\Return{\Var{m} & \Le & \Zero}{\False}
\Return{\Zero & \Le & \Suc[\Var{n}]}{\True}
\Return{\Suc[\Var{m}] & \Le & \Suc[\Var{n}]}{\Var{m} \Le \Var{n}}}
&
\Let{\Lookup & \PiTel{\Var{m}}{\Nat} & \PiTel{\Var{xs}}{\List{\Var{A}}}}{\Maybe{\Var{A}}}{
\Return{\Lookup & \Var{m} & \Nil}{\Nothing}
\Return{\Lookup & \Zero & (\Cons[\Var{a}\: \Var{xs}])}{\Just[\Var{a}]}
\Return{\Lookup & (\Suc[\Var{n}]) & (\Cons[\Var{a}\: \Var{xs}])}{\Lookup[\Var{n}\: \Var{xs}]}}
\end{array}
\]
}
\caption{Implementation of \(\_\Le\_\) and \(\Lookup\)}
\label{fig:le-lookup}
\end{figure}









This raises the question: what \emph{exactly} is the relation between
\(\_\Le\_\) and \(\Lookup\)? Also, could we use the implementation of
\(\_\Le\_\) to guide the construction of \(\Lookup\)? First, let us
work out the relation at the type level. To this end, we use ornaments
to explain how each individual datatype has been promoted when going
from \(\_\Le\_\) to \(\Lookup\):
\[
\begin{tikzpicture}
\matrix (m) [matrix of math nodes
            , row sep=3em
            , column sep=5em
            , text height=1.5ex
            , text depth=0.25ex
            , ampersand replacement=\&]
{ 
  \_\Le\_ \& \Nat \& \Nat           \& \Bool \\
  \Lookup \& \Nat \& \List{\Var{A}} \& \Maybe{\Var{A}} \\
};
\draw[double,->]
   (m-1-2) -- node[left] { \(\IdentityOrn{\Nat}\) } (m-2-2);
\draw[double,->]
   (m-1-3) -- node[left] { \(\OrnamentOf{\List{}}\) } (m-2-3);
\draw[double,->]
   (m-1-4) -- node[left] { \(\OrnamentOf{\Maybe{}}\) } (m-2-4);
\path 
   (m-1-1) -- (m-1-2) node [midway] {:}
   (m-1-2) -- (m-1-3) node [midway] {\(\To\)}
   (m-1-3) -- (m-1-4) node [midway] {\(\To\)}
   (m-2-1) -- (m-2-2) node [midway] {:}
   (m-2-2) -- (m-2-3) node [midway] {\(\To\)}
   (m-2-3) -- (m-2-4) node [midway] {\(\To\)};
\end{tikzpicture}
\]
Note that the first argument is ornamented to itself, or put
differently, it has been ornamented by the identity ornament.

Each of these ornaments come with a forgetful map, computed from the
ornamental algebra:
{
\[
\begin{array}{@{}c@{\:\:}c}
\LengthListDef 
&
\IsJustDef
\end{array}
\]
}
Using these forgetful map, the relation, at the computational level,
between \(\_\Le\_\) and \(\Lookup\) is uniquely established by
following the ornamentation of their types. This relation is naturally
expressed by the following \emph{coherence} property:
{
\[
\Forall{\Var{n}}{\Nat}
  \Forall{\Var{xs}}{\List{\Var{A}}}\:
    \IsJust[(\Lookup[\Var{n}\:\Var{xs}])] \PropEqual \Var{n} \Le \LengthList[\Var{xs}]
\]
}

Or, equivalently, using a commuting diagram:
\[
\begin{tikzpicture}
\matrix (m) [matrix of math nodes
            , row sep=3em
            , column sep=5em
            , text height=1.5ex
            , text depth=0.25ex
            , ampersand replacement=\&]
{ 
  \Nat \& \List{\Var{A}} \& \Maybe{\Var{A}} \\
  \Nat \& \Nat           \& \Bool \\
};
\path 
   (m-1-1) -- (m-1-2) node [midway] {\(\Times\)}
   (m-2-1) -- (m-2-2) node [midway] {\(\Times\)}
;
\draw[double,-]
   (m-1-1) -- (m-2-1)
;
\path[->]
   (m-2-2) edge node [below] {\(\_\Le\_\)} (m-2-3)
   (m-1-2) edge node [above] {\(\Lookup\)} (m-1-3)
   (m-1-2) edge node [left] {\(\LengthList\)} (m-2-2)
   (m-1-3) edge node [right] {\(\IsJust\)} (m-2-3)
;
\end{tikzpicture}
\]

Let us settle the vocabulary at this stage. We call the function we
start with the \emph{base function} (here, \(\_\Le\_\)), its type
being the \emph{base type} (here, \(\Nat \To \Nat \To \Bool\)). The
richer function type built by ornamenting the individual pieces is
called the \emph{functional ornament} (here, \(\Nat \To \List{A} \To
\Maybe{A}\)). A function inhabiting this type is called a
\emph{lifting} (here, \(\Lookup\)). A lifting is said to be
\emph{coherent} if it satisfies the coherence property. It is crucial
to understand that the coherence of a lifting is relative to a given
functional ornament: the same base function ornamented differently
would give rise to different coherence properties.


We now have a better grasp of the relation between the base function
and its lifting. However, \(\Lookup\) remains to be implemented while
making sure that it satisfies the coherence property. Traditionally,
one would stop here: one would implement \(\Lookup\) and prove the
coherence as a theorem. This works rather well in a system like
Coq~\citep{coq:manual} as it offers a powerful theorem proving
environment. It does not work so well in a system like
Agda~\citep{norell:agda} that does not offer tactics to its users,
forcing them to write explicit proof terms. It would not work at all
in Haskell with GADTs, as it does not have any theorem proving
capability.

However, we are not satisfied by this laborious approach:
if we have dependent types, why should we use them only for
\emph{proofs}, as an afterthought?
We should rather write a lookup function \emph{correct by
  construction}: by implementing a more precisely indexed version of
\(\Lookup\), the user can drive the index-level computations to
unfold, hence making the type-checker verify the necessary invariants.
We believe that this is how it should be: computers should
replace proofs by computation; humans should drive computers. The other way
around -- where humans are coerced into computing for computers
-- may seem surreal, yet it corresponds to the current situation in most
proof systems.


\Spacedcommand{\Ilookup}{\Function{ilookup}}
\newcommandx{\IMaybe}[2][2=\!]{\Canonical{IMaybe}\ifthenelse{\isempty{#1}}{}{_{#1}}\xspace\, #2}
\Spacedcommand{\ForgetIMaybe}{\Function{forgetIMaybe}}
\Spacedcommand{\MakeVec}{\Function{makeVec}}


To get the computer to work for us, we would rather implement the
function \(\Ilookup\):
{
\[
\Let{\Ilookup &
         \PiTel{\Var{m}}{\Nat} & 
         \PiTel{\Var{vs}}{\Vector{\Var{A}}[\Var{n}]}}
    {\IMaybe{\Var{A}}[(\Var{m} \Le \Var{n})]}{
\Return{\Ilookup & 
            \Var{m} & 
            \VNil}
       {\Nothing}
\Return{\Ilookup & \Zero & (\VCons[\Var{a}\:\Var{vs}])}{\Just[\Var{a}]}
\Return{\Ilookup & (\Suc[\Var{m}]) & (\VCons[\Var{a}\:\Var{vs}])}{\Ilookup[\Var{m}\: \Var{vs}]}}
\]
}
Where \(\IMaybe{A}\) is \(\Maybe{A}\) indexed by its truth as computed
by \(\IsJust\). It is defined as follows\footnote{Note that we have
  overloaded the constructors of \(\Maybe{}\) and \(\IMaybe{}\): for a
  bi-directional type-checker, there is no ambiguity as constructors
  are checked against their type.}:
{
\[
\Data{\IMaybe{}}{\Param{\Var{A}}{\Set} \Index{\Var{b}}{\Bool}}{\Set}{
 \Emit{\IMaybe{\Var{A}}}{\True}{\Just[\PiTel{\Var{a}}{\Var{A}}]}
 \Emit{\IMaybe{\Var{A}}}{\False}{\Nothing}}
\]
}
This comes with the following forgetful map\footnote{We depart
   slightly from the convention of calling \(\Constructor{refl}\) the
   inhabitant of the identity type \(a \PropEqual b\): instead we shall denote it
   \(\Refl{a}{b}\), hence being explicit about which equation is being
   proved.}:
{
\[
\Let{\ForgetIMaybe &
         \PiTel{\Var{mba}}{\IMaybe{\Var{A}}[\Var{b}]}}
    {\SigmaTimes{\Var{ma}}{\Maybe{\Var{A}}}
                {\IsJust[\Var{ma}] \PropEqual \Var{b}}}{
\Return{\ForgetIMaybe & (\Just[\Var{a}])}{\Pair{\Just[\Var{a}]}{\Refl{\True}{\True}}}
\Return{\ForgetIMaybe & \Nothing}{\Pair{\Nothing}{\Refl{\False}{\False}}}
}
\]
}

The rational behind \(\Ilookup\) is to \emph{index} the types of
\(\Lookup\) by their unornamented version, \ie the types of
\(\_\Le\_\). Hence, we can make sure that the result computed by
\(\Ilookup\) respects the output of \(\_\Le\_\) on the unornamented
indices: the result is correct \emph{by indexing}! The type of
\(\Ilookup\) is naturally derived from the ornamentation of
\(\_\Le\_\) into \(\Lookup\) and is uniquely determined by the
functional ornament we start with. Expounding further our vocabulary,
we call \emph{coherent liftings} these finely indexed functions that
are correct by construction.



\Spacedcommand{\CohLookup}{\Function{cohLookup}}


\citet{ko:modularising-inductive} use
ornaments to specify the coherence requirements
for functional liftings, but we work
the other way around, using ornaments to internalise coherence
requirements. From \(\Ilookup\), we can extract both \(\Lookup\) and its proof of
correctness \emph{without having written any proof term ourselves}:
{
\[
\Let{\Lookup & 
         \PiTel{\Var{m}}{\Nat} & 
         \PiTel{\Var{xs}}{\List{\Var{A}}}}
    {\Maybe{\Var{A}}}{
\Return{\Lookup & \Var{m} & \Var{xs}}
       {\Fst (\ForgetIMaybe[(\Ilookup[\Var{m}\: (\MakeVec[\Var{xs}])])])}}
\]
}
{
\[
\Let{\CohLookup & 
         \PiTel{\Var{n}}{\Nat} &
         \PiTel{\Var{xs}}{\List{\Var{A}}}}
    {\IsJust[(\Lookup[\Var{n}\:\Var{xs}])] \PropEqual \Var{n} \Le \LengthList[\Var{xs}]}{
\Return{\CohLookup & \Var{m} & \Var{xs}}
       {\Snd (\ForgetIMaybe[(\Ilookup[\Var{m}\: (\MakeVec[\Var{xs}])])])}}
\]
}
Where \(\MakeVec\) simply turns a list into a vector of the
corresponding length.



\Spacedcommand{\Vlookup}{\Function{vlookup}}

As a side comment, remark that the function \(\Ilookup\) is very
similar to the more familiar \(\Vlookup\) function:
\[
\Let{\Vlookup & \PiTel{\Var{m}}{\Fin[\Var{n}]} & \PiTel{\Var{vs}}{\Vector{\Var{A}}[\Var{n}]}}{\Id{\Var{A}}}{
\Impossible{\Vlookup & \Unreachable & \VNil}
\Return{\Vlookup & \FZero & (\VCons[\Var{a}\: \Var{xs}])}{\Var{a}}
\Return{\Vlookup & (\FSuc[\Var{n}]) & (\Cons[\Var{a}\: \Var{xs}])}{\Vlookup[\Var{n}\: \Var{xs}]}
}
\]
As we shall see later in Section~\ref{sec:remove-index-computation},
these two definitions are actually isomorphic, thanks to the following
equivalence:
\[
\PiTo{\Var{m}}{\Nat} \IMaybe{\Var{A}}{(\Var{m} \Le \Var{n})} \cong \Fin[n] \To \Var{A}
\]
Intuitively, we can move the constraint that \(m \Le n\) from the
result -- \ie \(\IMaybe{\Var{A}}{(\Var{m} \Le \Var{n})}\) -- to the
premises -- \ie \(\Fin[n]\). This matches our intuition that
\[
\Fin[\Var{n}] \cong \SigmaTimes{\Var{m}}{\Nat}{\Var{m} \Le \Var{n}}
\]
Hence, having implemented \(\Ilookup\), not only do we obtain
\(\Lookup\) and its coherence proof, but we also get the traditional
\(\Vlookup\) function.


With this example, we have manually unfolded the key steps of the
construction of a lifting of \(\_\Le\_\). Let us recapitulate each
steps:
\begin{itemize}
\item Start with a \emph{base function}, here \(\TypeAnn{\_\Le\_}{\Nat \To \Nat \To \Bool}\)
\item Ornament its inductive components as desired, here \(\Nat\) to
  \(\List{A}\) and \(\Bool\) to \(\Maybe{A}\) in order to describe the
  desired lifting, here \(\TypeAnn{\Lookup}{\Nat \To \List{\Var{A}} \To \Maybe{\Var{A}}}\) satisfying 
\(
\Forall{\Var{n}}{\Nat}
  \Forall{\Var{xs}}{\List{\Var{A}}}\:
    \IsJust[(\Lookup[\Var{n}\:\Var{xs}])] \PropEqual \Var{n} \Le \LengthList[\Var{xs}]
\)
\item Implement a carefully indexed version of the lifting, here
  \(\TypeAnn{\Ilookup}
            {\PiTel{\Var{m}}{\Nat}
              \PiTo{\Var{vs}}{\Vector{\Var{A}}[\Var{n}]} 
                \IMaybe{\Var{A}}[(\Var{m} \Le \Var{n})]}\)
\item Derive the lifting, here \(\Lookup\), and its coherence proof,
  without proving any theorem
\end{itemize}
Besides, the implementation of \(\Ilookup\) is not lost: this function
corresponds exactly to vector lookup, a function that one would have
implemented anyway.

This manual unfolding of the lifting is instructive: it involves a lot
of constructions on datatypes (here, the datatypes \(\List{A}\) and
\(\Maybe{A}\)) as well as on functions (here, the type of
\(\Ilookup\), the definition of \(\Lookup\) and its coherence
proof). Yet, it feels like a lot of these constructions could be
automated. In the next Section, we shall build the machinery to
describe these constructions and obtain them \emph{within} the type
theory itself.


\section{A universe of datatypes and their ornaments}
\label{sec:univ-data-orn}

In dependently typed systems such as Coq or Agda, datatypes are an
external entity: each datatype definition extends the type-theory
with new introduction and elimination forms. The validity of
datatypes is guaranteed by a positivity-checker that is part of the
meta-theory of the proof system. A consequence is that, from within
the type theory, it is not possible to create or manipulate datatype
definitions, as they belong to the meta-theory.


\subsection{A closed theory of datatypes}


In our previous work~\citep{dagand:levitation}, we have shown how to
internalise inductive families into type theory. The practical impact
of this approach is that we can manipulate datatype declarations as
first-class objects. We can program over datatype declarations and,
in particular, we can compute new datatypes from old. This is
particularly useful to formalise the notion of ornament entirely
within the type theory. This also has a theoretical impact: we do not
need to prove meta-theoretical properties of our constructions, we can
work in our type theory and use its logic as our formal system.

Note that our results are not restricted to this setting where
datatype definitions are internalised: all our constructions could be
justified at the meta-level and then be syntactically presented in a
language, such as, say, Agda or Haskell with GADTs. Working with an
internalised presentation, we can simply avoid these two levels of
logic and work in the logic provided by the type theory itself.


\renewcommand{\InterpretIDescDef}{
\Code{
  \InterpretIDesc{\PiTel{\Var{D}}{\IDesc[\Var{I}]}}\:
                  \PiTel{\Var{X}}{\Var{I} \To \Set} \::\:
      \Set \\
\begin{array}{@{}l@{}c@{}l}
\InterpretIDesc{\DVar[\Var{i}]}\:
    \Var{X} &
    \DoReturn &
    \Var{X}\: \Var{i} \\
\InterpretIDesc{\DUnit}\:
    \Var{X} &
    \DoReturn &
    \Unit \\
\InterpretIDesc{\DPi[\Var{S}\: \Var{T}]}\:
    \Var{X} &
    \DoReturn &
    \PiTo{\Var{s}}{\Var{S}}{\InterpretIDesc{\Var{T}\: \Var{s}}[\Var{X}]} \\
\InterpretIDesc{\DSigma[\Var{S}\: \Var{T}]}\:
    \Var{X} &
    \DoReturn &
    \SigmaTimes{\Var{s}}{\Var{S}}{\InterpretIDesc{\Var{T}\: \Var{s}}[\Var{X}]}\\
\end{array}
}}

\begin{figure}[t]
{
\[
\begin{array}{l|l}
\IDescDef 
&
\InterpretIDescDef
\end{array}
\]
}
\caption{Universe of inductive families}
\label{fig:universe-inductive-families}
\end{figure}

\newcommand{\collection}[1]{\blue{\left\{ \black{#1} \right\}}}
\newcommand{\collectionElim}[1]{\green{\left\{ \black{#1} \right\}}}

For the sake of completeness, let us recall a few definitions and
results from our previous work. As in previous work, our requirement
on the type theory are minimal: we will need \(\Sigma\)-,
\(\Pi\)-types, and at least two universes. For convenience, we require
a type of finite sets, which lets us build collections of
labels\footnote{We denote finite sets of tagged elements by
  \(\collection{\Tag{\CN{x}}, \Tag{\CN{y}, \Tag{\CN{z}},
      \ldots}}\). Their elimination principle consists of an
  exhaustive case enumeration and is denoted by
  \(\collectionElim{\Tag{\CN{x}} \mapsto vx, \Tag{\CN{y}} \mapsto vy,
    \Tag{\CN{z}} \mapsto vz, \ldots}\). If the tags are vertically
  aligned, we shall skip the separating comma.}. We also need a
pre-existing notion of propositional equality, upon which we make no
assumption. We internalise the inductive families by a universe
construction (Fig.~\ref{fig:universe-inductive-families}): an indexed
datatype is described by a function from its index to codes. The
codes are then interpreted to build the fix-point:
{
\[
\IMuDef
\]}
Inductive types are eliminated by a generic elimination
principle:
\[
\Code[ll]{
\TypeAnn{\Iinduction}
      {& \PiTel{\Var{D}}{\IDesc[\Var{I}]}
         \PiTel{\Var{P}}{\IMu[\Var{D}\: \Var{i}] \To \Set} \\
       & \PiTel{\Var{is}}{\PiTel{\Var{i}}{\Var{I}}
                          \PiTo{\Var{xs}}{\InterpretIDesc{\Var{D}\: \Var{i}}[(\IMu[\Var{D}])]}
                          \IAll[\Var{D}\: \Var{P}\: \Var{xs}] \To
                          \Var{P}\: (\In[\Var{xs}])
         } \\
       &
         \PiTel{\Var{x}}{\IMu[\Var{D}\: \Var{i}]} \To
         \Var{P}\: \Var{x}
        }}
\]
Where, intuitively, \(\IAll[\Var{D}\: \Var{P}\: \Var{xs}]\) enforces
that all sub-trees of \(\Var{xs}\) satisfy \(\Var{P}\): this
corresponds exactly to computing the inductive hypothesis necessary to
perform the induction step. In the categorical
literature~\citep{hermida:induction,
  atkey:inductive-refinement}, \(\IAll[\Var{D}]\) is 
denoted \(\hat{D}\).

For readability purposes, we use an informal notation to declare
datatypes. This notation is strongly inspired by Agda's datatype
declarations. Note that these definitions can always be turned into
\(\IDesc\) codes: when defining a datatype \(T\), we will denote
\(\IDescOf{T}\) the code it elaborates to. Similarly, we denote
\(\ElimOf{T}\) and \(\CaseOf{T}\) the induction principle and case
analysis operators associated with \(T\). These operations can be
implemented by generic programming, along the lines of
\citet{mcbride:constructions-on-constructors}. Formalising the
elaboration of datatypes definitions down to code is beyond the scope
of this paper. However, it is simple enough to be understood with a
few examples. Three key ideas are at play.

\paragraph{First, non-indexed datatypes definitions follow the ML tradition:}
we name the datatype and then comes a choice of constructors. For
example, \(\List{}\) would be written and elaborated as follows:
{
\[
\begin{array}{@{}c}
\ListDef \\
\rotatebox{-90}{\(\leadsto\)} \\
\begin{array}{@{}l@{}l}
\Lam{\Var{A}} 
\Lam{\Void} & \\
    & \DSigma[\collection{\begin{array}{l}\Tag{\CN{nil}}\\ \Tag{\CN{cons}}\end{array}}
             \collectionElim{\begin{array}{@{}l@{\:\:\mapsto\:}l}
                 \Tag{\CN{nil}}  & \DUnit\\ 
                 \Tag{\CN{cons}} & \DSigma[\Var{A}\: \Lam{\_}{\DVar[\Void]}]\end{array}}]
\end{array}
\end{array}
\]
}

\paragraph{Secondly, indexed datatypes can be defined following  the
Agda convention:} indices are \emph{constrained} to some particular
value. For example, \(\Vector{}\) could be defined by constraining the
index to be \(\Zero\) in the \(\VNil\) case and \(\Suc[n']\) for some
\(\TypeAnn{n'}{\Nat}\) in the \(\VCons\) case:
{\
\[
\begin{array}{c}
\VectorDefEquality
\\
\rotatebox{-90}{\(\leadsto\)} \\
\Lam{\Var{A}} 
\Lam{\Var{n}}
    {\begin{array}[t]{@{}l@{}l}
     \DSigma & \collection{
               \begin{array}{l}
                 \Tag{\CN{vnil}}\\
                 \Tag{\CN{vcons}}
               \end{array}} \\
             & \collectionElim{
               \begin{array}{@{}l@{\:\:\mapsto\:}l}  
                 \Tag{\CN{vnil}} &
                 \DSigma[(\Var{n} \PropEqual \Zero)\: \Lam{\_}{\DUnit}]\\
                 \Tag{\CN{vcons}} &
                 \begin{array}[t]{@{}l}
                 \DSigma[\Nat\: \Lam{\Var{n'}}
                   \DSigma[(\Var{n} \PropEqual \Suc[\Var{n'}]) \Lam{\_} \\
                   \quad\DSigma[\Var{A}\: \Lam{\_}
                   \DVar[\Var{n'}]]]]
                 \end{array}
               \end{array}}
      \end{array}}
\end{array}
\]
}
The elaboration naturally captures the constraints on indices by using
propositional equality. In the case of \(\Vector{}\), we abstract over
the index \(n\), introduce the choice of constructors with the
first \(\DSigma\) and, once constructors have been chosen, we restrict
\(n\) to its valid value(s): \(\Zero\) in the first case and
\(\Suc[n']\) for some \(n'\) in the second case.


\paragraph{Thirdly, we can compute over indices:} 
here, we make use of the crucial property that a datatype definition is
a \emph{function} from index to \(\IDesc\) codes. Hence, our notation
should reflect this ability to define datatypes as functions on their
index. For instance, inspired by \citet{brady:optimisations}, an
alternative presentation of vector would match on the index to
determine the constructor to be presented, hence removing the need for
constraints:
{
\[
\begin{array}{c}
\VectorDef \\
\rotatebox{-90}{\(\leadsto\)} \\
\begin{array}{l@{}l@{\:}l}
\Lam{\Var{n}} & & \\
    & \NatCase & \Var{n}\: (\Lam{\_}{\IDesc[\Nat]}) \\
    &          & \DUnit \\
    &          & (\Lam{\Var{n}}{\DSigma[\Var{A}\: \Lam{\_}{\DVar[\Var{n}]}]})
\end{array}
\end{array}
\]
}
In order to be fully explicit about computations, we use here the
Epigram~\citep{mcbride:view-from-left} \emph{by} (\(\DoBy\)) programming
gadget, which let us appeal to any elimination principle with a syntax
close to pattern-matching. However, standard pattern-matching
constructions~\citep{coquand:pattern-matching, norell:agda} would work
just as well. Again, we shall write pattern-matching definitions when
the recursion pattern is unremarkable.

Our syntax departs radically from the one adopted by Coq, Agda, and
GADTs in Haskell. It is crucial to understand that this is but
reflecting the actual semantics of inductive families: we can
\emph{compute} over indices, not merely constrain them to be what we
would like. With our syntax, we give the user the ability to write
these \emph{functions}: the reader should now understand a datatype
definition as a special kind of function definition, taking indices as
arguments, potentially computing over them, and eventually emitting a
choice of constructors.

\Spacedcommand{\Walk}{\Canonical{Walk}}
\Spacedcommand{\Up}{\Constructor{up}}
\Spacedcommand{\Down}{\Constructor{down}}
\newcommand{\Stop}{\Constructor{stop}}

As a final example, let us look at the datatype of infinite staircase
walks: at each step, we can unconditionally go \(\Up\). Besides, after
pattern-matching on the index, we can decide to \(\Stop\) on step
\(\Zero\) and, for any non-zero step, we can also go \(\Down\):
{
\[
\Data{\Walk}{\Index{\Var{n}}{\Nat}}{\Set}{
\Emit{\Walk}{\Var{n}}{\Up[\PiTel{\Var{w}}{\Walk[(\Suc[\Var{n}])]}]}
\Emit{\Walk}{\Zero}{\Stop}
\Emit{\Walk}{(\Suc[\Var{n}])}{\Down[\PiTel{\Var{w}}{\Walk[\Var{n}]}]}}
\]
}

\paragraph{Remark:} 
Note that we have been careful in using equality to \emph{introduce}
constraints here: our definition of datatypes is absolutely agnostic
in the notion of equality assumed by the underlying type theory. For
instance, our universe of inductive families cannot be used to
\emph{define} equality through the identity type: the identity type
would only \emph{expose} the underlying notion of equality to the
user. Concretely, the standard definition of the identity types is
presented and elaborated as follows:
{
\[
\newcommand{\Identity}{\Canonical{Id}}
\newcommand{\IdRefl}{\Constructor{refl}}
\begin{array}{@{}c}
\Data{\Identity}
     {\Param{\Var{a_1}}{\Var{A}} 
      \Index{\Var{a_2}}{\Var{A}}}
     {\Set}{
\Emit{\Identity}{\Var{a_1}\: \Constraint{\Var{a_2}}{\Var{a_1}}}{\IdRefl}
}
\\
\rotatebox{-90}{\(\leadsto\)}
\\
\begin{array}{@{}l}
\Lam{\Var{a_1}\: \Var{a_2}} \\
\quad \DSigma[\collection{\Tag{\CN{refl}}}
              \collectionElim{\DSigma[(\Var{a_2} \PropEqual \Var{a_1}) 
                                      \Lam{\_}{\DUnit}]}]
\end{array}
\end{array}
\]
}
We have been careful to maintain our presentation of ornaments and
functional ornaments similarly equality agnostic.


\subsection{Ornaments}


Originally, \citet{mcbride:ornament} presented the notion of ornament for a
universe where the indices a
constructor targets could be enforced \emph{only} by equality
constraints. As a consequence, in that simpler setting, computing types
from indices was impossible. We shall now adapt the original
definition to our setting.

Just as the original definition, an ornament is defined upon a base
datatype -- specified by a function \(\TypeAnn{D}{I \To \IDesc[I]}\)
-- and indices are refined up to a reindexing function
\(\TypeAnn{re}{J \To I}\). The difference in our setting is that, just
as the code of datatypes can be computed from the indices, we want
the ornament to be computable from its \(J\)-index. Hence, an ornament
is a function from \(\TypeAnn{j}{J}\) to ornament codes
describing the ornamentation of \(D\: (re\: j)\):
{
\[
\ornDef
\]
}
As for the ornament codes themselves, they are similar to the original
definition: we shall be able to \emph{copy} the base datatype,
\emph{extend} it by inserting sets, or \emph{refine} the indexing
subject to the relation imposed by \(re\). However, we also have the
\(J\)-index in our context: following Brady's insight that
\emph{inductive families need not store their
  indices}~\citep{brady:optimisations}, we could as well \emph{delete}
parts of a datatype definition as long as we can recover the
suppressed bits from the index. Hence, we obtain the following
code\footnote{The inverse image of a function is defined by:
{
\[
\InverseDef
\]
}}:
{
\[
\OrnDef
\]
}


Given an ornament, we can interpret it as the datatype it
describes. The implementation consists in traversing the ornament
code, introducing a \(\DSigma\) when inserting new data and computing
the ornament at the replaced value when deleting some redundant data:
{
\[
\Code{
\InterpretornDef \\
\begin{array}{@{\quad}ll}
\where &  \\
       & \InterpretOrnDef
\end{array}}
\]
}
Note that in the \(\ODelete\) case, no \(\DSigma\) code is generated:
the set \(S\) has been deleted from the original datatype. The
witness of this existential is instead provided by \(\CN{replace}\).


Once again, we adopt an informal notation to describe ornaments
conveniently. The idea is to simply mirror our \(\Kwd{data}\)
definition, adding \(\Kwd{from}\) which datatype the ornament is
defined. When specifying
a constructor, we can then extend it with a new element using
\(\OrnInsert{\Var{s}}{S}\) or delete an element originally named
\(\Var{s}\) by giving its value with
\(\OrnDelete{\Var{s}}{\CN{value}}\). Some typical examples of
extension are presented in Figure~\ref{fig:example-ornament-syntax}.

\Spacedcommand{\FinOpt}{\Canonical{Fin'}}

\begin{figure}[tb]
{
\[
\Code{
\ListOrnDef \\
\\
\VectorOrnDefEquality \\
\\
\FinOrnDef
}
\]
}
\caption{Examples of ornament}
\label{fig:example-ornament-syntax}
\end{figure}

\renewcommand{\VectorOrnDef}{
  \Ornament{\Vector{}}
           {\Param{\Var{A}}{\Set}
            \Index{\Var{n}}{\Nat}}
           {\List{\Var{A}} \if 0 ((\Lam{\_}{\Void})\: \Var{n}) \fi}
           {
\Emit{\Vector{\Var{A}}}
     {\Zero}
     {\VNil}
\Emit{\Vector{\Var{A}}}
     {(\Suc[\Var{n'}])}
     {\VCons[\PiTel{\Var{a}}{\Var{A}}
             \PiTel{\Var{vs}}{\Vector{\Var{A}}{\Var{n'}}}]}
}
}

While the definition \(\Vector{}\) in
Figure~\ref{fig:example-ornament-syntax} mirrors Agda's convention
of constraining indices with equality, our definition of ornaments
lets us define a version of \(\Vector{}\) that does not store its
indices:
{
\[
\VectorOrnDef
\]
}
Note that such a definition was unavailable in the basic
presentation~\citep{mcbride:ornament}. \citet{brady:optimisations} call this operation
\emph{detagging}: the constructors of the datatype are
determined by the index. The definition of \(\Fin\) given in
Figure~\ref{fig:example-ornament-syntax} is also subject to an
optimisation: by matching the index, we can avoid the duplication of
\(n\) by deleting \(n'\) with the matched predecessor and deleting the
resulting, obvious proof. Hence, \(\Fin\) can be further ornamented to
the optimised \(\FinOpt\), which makes crucial use of deletion:
{
\[
\Ornament{\FinOpt}
         {\Index{\Var{n}}{\Nat}}
         {\Fin \if 0 (\Function{id}\: \Var{n}) \fi}{
\Emit{\FinOpt}{\Zero}{\OrnInsert{\Var{b}}{\Empty} (\EmptyElim[\Var{b}]) 
                          \qquad\qquad\CommentLine{no constructor}}
\Emit{\FinOpt}{(\Suc[\Var{n}])}{\FZero\: \OrnDelete{\Var{n'}}{\Var{n}}
                                          \OrnDelete{\Var{q}}{\Refl{\Var{n}}{\Var{n}}}}
                     \OrEmit{\FSuc[\OrnDelete{\Var{n'}}{\Var{n}}
                                         \OrnDelete{\Var{q}}{\Refl{\Var{n}}{\Var{n}}}
                                         \PiTel{\Var{fn}}{\FinOpt[\Var{n'}]}]}
}
\]
}
Again, this definition was previously unavailable to us. Besides, we
are making crucial use of the deletion
ornament to avoid duplication.
\citet{brady:optimisations} call this operation \emph{forcing}:
the content of the constructors, here \(n'\) and the constraint, are
computed from the index.

Just as the datatype declaration syntax was elaborated to \(\IDesc\)
codes, this high-level syntax is elaborated to ornament codes. The
formal description of this translation is beyond the scope of this
paper. From the definition of an ornamented type \(T\), we will assume
the existence of its corresponding ornament code \(\OrnamentOf{T}\).

As described by \citet{mcbride:ornament}, every ornament induces an
\emph{ornamental algebra}: intuitively, an algebra that forgets the
extra data, hence mapping the ornamented datatype back to its
unornamented form. From an ornament
\(\TypeAnn{O}{\orn[re\: D]}\), there is a natural
transformation from the ornamented functor down to the unornamented
one, which we denote:
{
\[
\TypeAnn{\OrnForgetNat{O}}
        {\PiTel{\Var{X}}{\Var{I} \To \Set}
         \PiTo{\Var{j}}{\Var{J}}
         \InterpretIDesc{\Interpretorn{\Var{O}}\: \Var{j}}[(\Var{X} \Compose \Var{re})] \To
         \InterpretIDesc{\Var{D}\: (\Var{re}\: \Var{j})}[\Var{X}]}
\]}
Applied with \(\IMu[D]\) for \(X\) and post-composed with
\(\In\), this natural transformation induces the ornamental algebra:
{
\[
\TypeAnn{\OrnForgetAlg{O}}
        {\PiTo{\Var{j}}{\Var{J}}
         \InterpretIDesc{\Interpretorn{\Var{O}}\: \Var{j}}[(\IMu[\Var{D}] \Compose \Var{re})] \To
         \IMu[\Var{D}\: (\Var{re}\: \Var{j})]}
\]
}
In turn, this algebra induces an ornamental forgetful map denoted:
{
\[
  \TypeAnn{\OrnForget{O}}
          {\PiTo{\Var{j}}{\Var{J}} 
            \IMu[\Interpretorn{\Var{O}}\: \Var{j}] \To
            \IMu[\Var{D}\: (\Var{re}\: \Var{j})]}
\]
}
We do not re-implement these functions here: it is straightforward to
update the original definitions to our setting.

\subsubsection{Algebraic ornaments}
\label{sec:algebraic-ornament}


\newcommand{\Fold}[1]{\green{\llparenthesis} #1 \green{\rrparenthesis}\xspace}

An important class of datatypes is constructed by \emph{algebraic
  ornamentation} over a base datatype. The idea of an algebraic
ornament is to index an inductive type by the result of a fold over
the original data: from the code \(\TypeAnn{D}{I \To \IDesc[I]}\) and
an algebra \(\TypeAnn{\alpha}{\PiTo{\Var{i}}{I} \InterpretIDesc{D\:
    \Var{i}}[X] \To X\:\Var{i}}\), there is an ornament that defines a
code \(\TypeAnn{\OrnAlgebraic{D}{\alpha}}{\SigmaTimes{\Var{i}}{I}{X\:
    \Var{i}} \To \IDesc[\SigmaTimes{\Var{i}}{I}{X\: \Var{i}}]}\) with
the property that:
{
\[
\IMu[\OrnAlgebraic{D}{\alpha}\: (i , x)] 
  \cong 
\SigmaTimes{\Var{t}}{\IMu[D\: i]}{\Fold{\alpha}\: \Var{t} \PropEqual x}
\]
}
We shall indiscriminately use \(\OrnAlgebraic{D}{\alpha}\) to refer to
the ornament and the resulting datatype. Seen as a refinement type, the
correctness property states that \(\IMu[\OrnAlgebraic{D}{\alpha}\: (i
  , x)] \cong \{ \Var{t} \in \IMu[D\: i] \:|\: \Fold{\alpha}\: \Var{t} = x\}\). The
type theoretic construction of \(\OrnAlgebraic{D}{\alpha}\) is
described by \citet{mcbride:ornament}. We shall not reiterate it here,
the implementation being essentially the same for our universe. A
categorical presentation was also given in
\citet{atkey:inductive-refinement} in which the connection with
refinement types was explored.

Constructively, the correctness property gives us two (mutually
inverse) functions. The direction {\(\IMu[\OrnAlgebraic{D}{\alpha} (i , x)] \To
\SigmaTimes{\Var{t}}{\IMu[D\: i]}{\Fold{\alpha}\: \Var{t} \PropEqual x}\)}
relies on the generic \(\OrnForget{\OrnAlgebraic{D}{\alpha}}\) function to
compute the first component of the pair and gives us the following
theorem:
{
\[
\TypeAnn{\OAAO}{
      \Forall{\Var{t^\alpha}}{\IMu[\OrnAlgebraic{D}{\alpha} (\Var{i}, \Var{x})]}\:
        \Fold{\alpha}\: (\OrnForget{\OrnAlgebraic{D}{\alpha}}\: \Var{t^\alpha})
            \PropEqual 
        \Var{x}}
\]
}
This corresponds to the \(\Function{Recomputation}\) theorem of
\citet{mcbride:ornament}. We shall not reprove it here, the
construction being similar. In the other direction, the isomorphism
gives us a function of type:
{
\[
\SigmaTimes{\Var{t}}
           {\IMu[D\: i]}
           {\Fold{\alpha}\: \Var{t} \PropEqual x} \To
               \IMu[\OrnAlgebraic{D}{\alpha} (i , x)]
\] 
}
Put in full and simplifying the equation, this corresponds to the
following function:
{
\[
\TypeAnn{\OrnMake{D}{\alpha}}
        {\PiTo{\Var{t}}{\IMu[D\: \Var{i}]}
         \IMu[\OrnAlgebraic{D}{\alpha} (\Var{i} , \Fold{\alpha}\: \Var{t})]}
\]
}
This corresponds to the \(\Function{remember}\) function of
\citet{mcbride:ornament}. Again, we will assume this construction
here.


\subsubsection{Reornaments}
\label{sec:reornament}


In this paper, we are interested in a special sub-class of algebraic
ornaments. As we have seen, every ornament \(O\) induces an ornamental
algebra \(\OrnForgetAlg{O}\), which forgets the extra information
introduced by the ornament. Hence, given a datatype \(D\) and an
ornament \(O_D\) of \(D\), we can algebraically ornament
\(\Interpretorn{O_D}\) using the ornamental algebra
\(\OrnForgetAlg{O_D}\). We denote the resulting ornament
\(\OrnAlgOrn{D}{O_D}\).


\renewcommand{\VectorDef}{
\Data{\Vector{}}
     {\Param{\Var{A}}{\Set}
      \Index{\Var{n}}{\Nat}}
     {\Set}{
\Emit{\Vector{\Var{A}}}{\Zero}{\VNil}
\Emit{\Vector{\Var{A}}}
     {(\Suc[\Var{n}])}
     {\VCons[\PiTel{\Var{a}}{\Var{A}}
             \PiTel{\Var{vs}}{\Vector{\Var{A}}[\Var{n}]}]}
}}

\citet{mcbride:ornament} calls this object the \emph{algebraic
  ornament by the ornamental algebra}. For brevity, we call the object
\(\OrnAlgOrn{D}{O_D}\) the \emph{reornament} of \(O_D\). Again, we
shall overload \(\OrnAlgOrn{D}{O_D}\) to denote both the ornament and
the resulting datatype. A standard example of reornament is
\(\Vector{}\): it is the reornament of \(\OrnamentOf{\List{}}\). Put
otherwise, a vector is the algebraic ornament of \(\List{}\) by the
algebra computing its length, \ie the ornamental algebra from
\(\List{}\) to \(\Nat\).

\Spacedcommand{\ReornExtension}{\Function{Extension}}
\Spacedcommand{\ReornStructure}{\Function{Structure}}

Reornaments can be implemented straightforwardly by unfolding their
definition: first, compute the ornamental algebra and, second,
construct the algebraic ornament by this algebra. However, such a
simplistic construction introduces a lot of spurious equality
constraints and duplication of information. For instance, using this
naive definition of reornaments, a vector indexed by \(n\) is
constructed as \emph{any} list \emph{as long as} it is of length
\(n\).

We can adopt a more fine-grained approach yielding an isomorphic but
better structured datatype. In our setting, where we can compute over
the index, a finer construction of the \(\Vector{}\) reornament would
be as follows:
\begin{itemize}
  \item We retrieve the index, hence obtaining \(n\)~;
  \item By inspecting the ornament \(\OrnamentOf{\List{}}\), we obtain
    \emph{exactly} the information by which \(n\) is \emph{extended}
    into a list: if \(n = 0\), no supplementary information is needed
    and if \(n = \Suc[n']\), we need to extend it with an
    \(\TypeAnn{a}{A}\). We call this the \(\ReornExtension\) of
    \(n\)~;
  \item By inspecting the ornament \(\OrnamentOf{\List{}}\) again, we
    obtain the recursive structure of the reornament by
    \emph{deleting} the data already fully determined by the index and
    its extension, and \emph{refining} the indexing discipline: the
    tail of a vector of size \(\Suc[n']\) is a vector of size
    \(n'\). The recursive structure is denoted by \(\ReornStructure\).
\end{itemize}

Let us formalise this intuition for any ornament. By the coherence
property, we know that for any index \(\TypeAnn{t}{\IMu[D]}\), the
reornament \(\TypeAnn{t^{++}}{\IMu[\OrnAlgOrn{D}{O_D}\: t]}\) is
isomorphic to the comprehension
\(
\{ \TypeAnn{\Var{t^+}}{\IMu[\Interpretorn{O}\: j]}
        \:|\: \OrnForget{O}\: \Var{t^+} = t \}
\).
Note that the equality constraints are introduced only to ensure
that \(t^+\) is built from \(t\) through the ornament. Now, in our
setting, we could enforce this constraint \emph{by construction}: from
the ornament \(O_D\) and \(t\), we can compute the set of valid
extensions of \(t\) giving a \(t^+\) such that forgetting the
extension gives \(t\) back:
\[
\Code{
\Let{\ReornExtension &
         \PiTel{\Var{O}}{\Orn[\Var{re}\: \Var{j}\: \Var{D}]} &
         \PiTel{\Var{xs}}{\InterpretIDesc{\Var{D}}[\IMu[\Var{D}]]}}
    {\Set}{
\Return{\ReornExtension &
        (\DVar{(\Inv[\Var{j}])}) &
        t}
       {\Unit}
\Return{\ReornExtension &
        \DUnit &
        \Void}
       {\Unit}
\Return{\ReornExtension &
        (\DPi[\Var{T^+}]) &
        \Var{f}}
       {\PiTo{\Var{s}}{\Var{S}}{\ReornExtension[(\Var{T^+}\: \Var{s})\: (\Var{f}\: \Var{s})]}}
\Return{\ReornExtension &
        (\DSigma[\Var{T^+}]) &
        \Pair{\Var{s}}{\Var{xs}}}
       {\ReornExtension[(\Var{T^+}\: \Var{s})\: \Var{xs}]}
\Return{\ReornExtension &
        (\OInsert[\Var{S}\: \Var{D^+}]) &
        \Var{xs}}
       {\SigmaTimes{\Var{s}}{\Var{S}}{\ReornExtension[(\Var{D^+}\: \Var{s})\: \Var{xs}]}}
\Return{\ReornExtension & 
        (\ODelete[\Var{\CN{replace}}\: \Var{T^+}]) &
        \Pair{\Var{s}}{\Var{xs}}}
       {\SigmaTimes{\Var{q}}{\Var{s} \PropEqual \Var{\CN{replace}}\: \Var{j}}
                   {\ReornExtension[\Var{T^+}\: \Var{xs}]}} 
}}
\]
The next step consists in building the recursive structure of the
reornamented type. Again, the recursive structure is entirely
described by the ornament together with the index \(t\): the ornament
gives the recursive structure of \(t^+\) while the index \(t\) specify
the indexing strategy of the sub-nodes: sub-nodes of \(t^{++}\) must be
indexed by the corresponding sub-nodes of \(t\). Hence, we obtain the
recursive structure of \(t^{++}\) by traversing the ornament
definition while unfolding \(t\) along the way in order to reach its
sub-nodes. On a variable, we index by the value specified by the
ornament and the \(t\) sub-node we have reached. On a \(\DSigma\) and
\(\OInsert\), we can delete them to avoid information duplication: the
information is already provided by the index in the case of
\(\DSigma\) and by the extension in the case of \(\OInsert\). The
formal definition is as follows:
\[
\Code{
\ReornStructure\:
     \PiTel{\Var{O}}{\Orn[\Var{re}\: \Var{j}\: \Var{D}]}
     \PiTel{\Var{xs}}{\InterpretIDesc{\Var{D}}[\IMu[\Var{D}]]}
     \PiTel{\Var{e}}{\ReornExtension[\Var{O}\: \Var{xs}]}
  \::\:  \Orn[\Fst\: \Pair{\Var{j}}{\Var{t}}\: \Interpretorn{\Var{O}}] \\
\begin{array}{@{}l@{\:}c@{}c@{}c@{}c@{}l}
\ReornStructure &
        (\DVar[(\Inv[\Var{j}])]) &
        \Var{t} &
        \Void &
        \DoReturn &
        \DVar[(\Inv[\Pair{\Var{j}}{\Var{t}}])] \\
\ReornStructure &
        \DUnit &
        \Void &
        \Void &
        \DoReturn &
        \DUnit \\
\ReornStructure &
        (\DPi[\Var{T^+}]) &
        \Var{f} &
        \Var{e} &
        \DoReturn &
        \DPi[\Lam{\Var{s}}{\ReornStructure[(\Var{T^+}\: \Var{s})\: (\Var{f}\: \Var{s})\: (\Var{e}\: \Var{s})]}] \\
\ReornStructure & 
        (\DSigma[\Var{T^+}]) &
        \Pair{\Var{s}}{\Var{xs}} &
        \Var{e} &
        \DoReturn &
        \ODelete[\Var{s}\: (\ReornStructure[(\Var{T^+}\: \Var{s})\: \Var{xs}\: \Var{e}])] \\
\ReornStructure &
        (\OInsert[\Var{S}\: \Var{D^+}]) &
        \Var{xs} &
        \Pair{\Var{s}}{\Var{e}} &
        \DoReturn &
        \ODelete[\Var{s}\: (\ReornStructure[(\Var{D^+}\: \Var{s})\: \Var{xs}\: \Var{e}])] \\
\end{array}\\
\ReornStructure\:
        (\ODelete[\Var{\CN{replace}}\: \Var{T^+}])\:
        \Pair{\Var{\CN{replace}}}{\Var{xs}}\:
        \Pair{\Refl{\Var{s}}{\Var{\CN{replace}}}}{\Var{e}}
        \DoReturn
        \ReornStructure[\Var{T^+}\: \Var{xs}\: \Var{e}]
}
\]
A reornament is thus the \(\ReornExtension\) of its index followed by
the recursive structure as defined by \(\ReornStructure\). Thus, we
define the associated reornament at index \(\TypeAnn{t =
  \In[xs]}{\IMu[D]}\) by, first, inserting the valid extensions of
\(t\) with \(\ReornExtension\), then, building the recursive structure
using \(\ReornStructure\):
{
\[
\Let{\Reornament & \PiTel{\Var{O}}{\orn[\Var{re}\: \Var{D}]}}
    {\orn[\Fst\: \Interpretorn{\Var{O}}]}{
\multicolumn{5}{@{}l}{
\Reornament \: \Var{O} \DoReturn
       \begin{array}[t]{@{}l@{}l}
           \Lam{\Pair{\Var{j}}
                  {\In[\Var{xs}]}} & 
              \OInsert\: (\ReornExtension[(\Var{O}\: \Var{j})\: \Var{xs}])\: \Lam{\Var{e}} \\
            & \ReornStructure[(\Var{O}\: \Var{j})\: \Var{xs}\: \Var{e}]
         \end{array}}}
\]
}
Applied to the reonarment of \(\OrnamentOf{\List{}}\),
this construction gives the fully Brady-optimised -- detagged and
forced -- version of \(\Vector{}\), here written in full:
{
\[
\VectorDef
\]
}

Note that our ability to \emph{compute} over the index is crucial for
this construction to work. Also, it is isomorphic to the datatype one
would have obtained with the algebraic ornament of the ornamental
algebra. Consequently, the correctness property of algebraic ornaments
is still valid here: constructively, we get the \(\OAAO\) theorem in
one direction and the \(\OrnMake{*}{}\) function in the other
direction.

In this Section, we have adapted the notion of ornament to our
universe of datatypes. In doing so, we have introduced the concept of
a deletion ornament, using the indexing to remove duplicated
information in the datatypes. This has proved useful to simplify the
definition of reornaments. We shall see how this can be turned to
our advantage when we transport functions across ornaments.


\section{A universe of functions and their ornaments}
\label{sec:lift}


We are now going to generalise the notion of ornament to functions. In
order to do this, we first need to be able, in type theory, to
manipulate functions and especially their types. Hence, we define a
universe of functions. With it, we will be able to write generic
programs over the class of functions captured by our universe.

Using this technology, we define a functional ornament as a decoration
over the universe of functions. The liftings implementing the
functional ornament are related to the base function by a coherence
property. To minimise the theorem proving burden induced by coherence
proofs, we expand our system with \emph{patches}: a patch is the type
of the functions that satisfy the coherence property \emph{by
  construction}. Finally, and still writing generic programs, we show
how we can automatically project the lifting and its coherence
certificate out of a patch.


\subsection{A universe of functions}

\newcommand{\TypeUniv}{\Canonical{Type}}
\newcommand{\muTo}[2]{\red{\mu\!\{} #1 \, #2 \mathop{\red{\}\!\!\!\rightarrow}}}
\newcommand{\muTimes}[2]{\red{\mu\!\{} #1 \, #2 \mathop{\red{\}\!\times}}}
\newcommand{\unit}{\Constructor{1}}

\Spacedcommand{\CaseType}{\Function{Case-Type}}

\newcommand{\interpretType}[1]{\green{\llbracket}#1\green{\rrbracket_{\CN{Type}}}}


For clarity of exposition, we restrict our language of types to the
bare minimum: a type can either be an exponential which domain is an
inductive object, or a product which first component is an inductive
object, or the unit type -- used as a termination symbol:
{
\[\Code{
\Data{\TypeUniv}{}{\Set[1]}{
\Emit{\TypeUniv}{}
     {\muTo{\PiTel{\Var{D}}{\Var{I} \To \IDesc[\Var{I}]}}
           {\PiTel{\Var{i}}{\Var{I}}}
           {\PiTel{\Var{T}}{\TypeUniv}}}
\OrEmit{\muTimes{\PiTel{\Var{D}}{\Var{I} \To \IDesc[\Var{I}]}}
                  {\PiTel{\Var{i}}{\Var{I}}}
                  {\PiTel{\Var{T}}{\TypeUniv}}}
\OrEmit{\unit}
}
}\]
}
Hence, this universe codes the function space from some (maybe none)
inductive types to some (maybe none) inductive types. Concretely, the
codes are interpreted as follows:
{
\[
\Let{\interpretType{\PiTel{\Var{T}}{\TypeUniv}}}{\Set}{
\Return{\interpretType{\muTo{\Var{D}}{\Var{i}}{\Var{T}}}}
       {\IMu[\Var{D}\: \Var{i}] \To \interpretType{\Var{T}}}
\Return{\interpretType{\muTimes{\Var{D}}{\Var{i}}{\Var{T}}}}
       {\IMu[\Var{D}\: \Var{i}] \Times \interpretType{\Var{T}}}
\Return{\interpretType{\unit}}
       {\Unit}
}
\]
}

The constructions we develop below could be extended to a more
powerful universe -- such as one supporting non-inductive sets or
having dependent functions and pairs. However, this would needlessly
complicate our exposition.



\newcommand{\LeType}{\Function{type}\green{<}}
\newcommand{\LeExplicitDef}{
\Let{\_ & \Le & \_}{\interpretType{\LeType}}{
\By{\Var{m} & \Le & \Var{n}}{\NatCase[\Var{n}]}
    \Return{\quad \Var{m} & \Le & \Zero}
           {\Pair{\False}{\Void}}
    \By{\quad \Var{m} & \Le & \Suc[\Var{n}]}
       {\NatElim[\Var{m}]}
    \Return{\quad\quad \Zero & \Le & \Suc[\Var{n}]}
           {\Pair{\True}{\Void}}
    \Return{\quad\quad \Suc[\Var{m}] & \Le & \Suc[\Var{n}]}
           {\Var{m} \Le \Var{n}}
}}

\begin{example}[Coding \(\_\Le\_\)]

Written in the universe of function types, the type of \(\_\Le\_\) is:
{
\[
\Let{\LeType}{\TypeUniv}{
\Return{\LeType}{\muTo{\IDescOf{\Nat}}{\Void}
                      {\muTo{\IDescOf{\Nat}}{\Void}
                            {\muTimes{\IDescOf{\Bool}}{\Void}
                                     {\unit}}}}
}
\]
}
The implementation of \(\_\Le\_\) is essentially the same as
earlier, excepted that it must now return a pair of a boolean and an
inhabitant of the unit type. To be explicit about the recursion
pattern of this function, we make use of Epigram's \emph{by} (\(\DoBy\))
construct~\citep{mcbride:view-from-left}:
{
\[
\LeExplicitDef
\]
}
That is to say: we first do a case analysis on \(n\) and then, in the successor
case, we proceed by induction over \(m\).

\end{example}

\newcommand{\PlusType}{\Function{type}\green{+}}
\newcommand{\PlusExplicitDef}{
\Let{\_ & \Plus & \_}{\interpretType{\PlusType}}{
\By{\Var{m} & \Plus & \Var{n}}{\NatCase[\Var{m}]}
    \Return{\quad \Zero & \Plus & \Var{n}}
           {\Pair{\Var{n}}{\Void}}
    \Return{\quad \Suc[\Var{m}] & \Plus & \Var{n}}
           {\Pair{\Suc[\Var{m} \Plus \Var{n}]}{\Void}}
}}

\begin{example}[Coding \(\_\Plus\_\)]

Written in the universe of function types, the type of \(\_\Plus\_\) is:
{
\[
\Let{\PlusType}{\TypeUniv}{
\Return{\PlusType}{\muTo{\IDescOf{\Nat}}{\Void}
                      {\muTo{\IDescOf{\Nat}}{\Void}
                            {\muTimes{\IDescOf{\Nat}}{\Void}
                                     {\unit}}}}
}
\]
}
Again, up to a multiplication by \(\Unit\), the implementation of
\(\_\Plus\_\) is left unchanged:
{
\[
\PlusExplicitDef
\]
}

\end{example}


\subsection{Functional ornament}


\Spacedcommand{\FunOrn}{\Canonical{FunOrn}}
\newcommand{\muOrnTo}[2]{\red{\mu^+\!\{} #1 \, #2 \mathop{\red{\}\!\!\!\rightarrow}}}
\newcommand{\muOrnTimes}[2]{\red{\mu^+\!\{} #1 \, #2 \mathop{\red{\}\!\times}}}

\newcommand{\interpretFunOrn}[1]{\green{\llbracket}#1\green{\rrbracket_{\CN{FunOrn}}}}

\Spacedcommand{\Coherence}{\Function{Coherence}}

From the universe of function types, it is now straightforward to
define the notion of functional ornament: we traverse the type and
ornament the inductive types as we go. Note that it is always possible
to leave an object unornamented: we ornament by the identity that
simply copies the original definition. Hence, we obtain the following
definition:
\[
\Data{\FunOrn}
     {\Index{\Var{T}}{\TypeUniv}}
     {\Set[1]}{
  \Emit{\:\:\FunOrn}
       {(\muTo{\Var{D}}{\Var{i}}{\Var{T}})}
       {\muOrnTo{\PiTel{\Var{O}}{\orn[\Var{re}\: \Var{D}]}}
                {\PiTel{\Var{j}}{\Var{re} \Inverse \Var{i}}}
                {\PiTel{\Var{T^+}}{\FunOrn[\Var{T}]}}}
  \Emit{\:\:\FunOrn}
       {(\muTimes{\Var{D}}{\Var{i}}{\Var{T}})}
       {\muOrnTimes{\PiTel{\Var{O}}{\orn[\Var{re}\: \Var{D}]}}
                   {\PiTel{\Var{j}}{\Var{re} \Inverse \Var{i}}}
                   {\PiTel{\Var{T^+}}{\FunOrn[\Var{T}]}}}
  \Emit{\:\:\FunOrn}
       {\unit}
       {\unit}
}
\]
From a functional ornament, we get the type of the liftings by
interpreting the ornaments as we go along:
\[
\Let{\interpretFunOrn{\PiTel{\Var{T^+}}{\FunOrn\: \Var{T}}}}{\Set}{
\Return{\interpretFunOrn{\muOrnTo{\Var{O}}{(\Inv[\Var{j}])}{\Var{T^+}}}}
       {\IMu[\Interpretorn{\Var{O}}\: \Var{j}] \To \interpretFunOrn{\Var{T^+}}}
\Return{\interpretFunOrn{\muOrnTimes{\Var{O}}{(\Inv[\Var{j}])}{\Var{T^+}}}}
       {\IMu[\Interpretorn{\Var{O}}\: \Var{j}] \Times \interpretFunOrn{\Var{T^+}}}
\Return{\interpretFunOrn{\unit}}
       {\Unit}
}
\]
We will want our ornamented function to be \emph{coherent} with the
base function we started with: for a function \(\TypeAnn{f}{\IMu[D]
  \To \IMu[E]}\), the ornamented function
\(\TypeAnn{f^+}{\IMu[\Interpretorn{O_D}] \To
  \IMu[\Interpretorn{O_E}]}\) is said to be coherent with \(f\) if the following diagram commutes:
\[
\begin{tikzpicture}
\matrix (m) [matrix of math nodes
            , row sep=3em
            , column sep=5em
            , text height=1.5ex
            , text depth=0.25ex
            , ampersand replacement=\&]
{ 
   \IMu[\Interpretorn{O_D}] \& \IMu[\Interpretorn{O_E}]     \\
   \IMu[D]                  \& \IMu[E]                      \\
};
\path[->] 
   (m-1-1) edge node[left] { \(\OrnForget{O_D}\) } (m-2-1)
   (m-1-2) edge node[right] { \(\OrnForget{O_E}\) } (m-2-2)
   (m-2-1) edge node[below] { \(f\) } (m-2-2)
   (m-1-1) edge node[above] { \(f^+\) } (m-1-2);
\end{tikzpicture}
\]
Or, equivalently in type theory:
{
\[
\Forall{\Var{x^+}}{\IMu[\Interpretorn{O_D}]}
    f \: (\OrnForget{O_D}\: \Var{x^+}) 
      \PropEqual 
    \OrnForget{O_E}\: (f^+\: \Var{x^+})
\]
}
To generalise the definition of coherence to any arity, we generically
define it by induction over the universe of functional ornaments:
{
\[
\Code{
\Coherence
         \PiTel{\Var{T^+}}{\FunOrn[\Var{T}]} 
         \PiTel{\Var{f}}{\interpretType{\Var{T}}} 
         \PiTel{\Var{f^+}}{\interpretFunOrn{\Var{T^+}}}
       : \Set \\
\begin{array}{@{}l@{\:}c@{\:}c@{\:}c@{}c@{}l}
\Coherence &
            (\muOrnTo{\Var{O}}{(\Inv[\Var{j}])}{\Var{T^+}}) &
            \Var{f} &
            \Var{f^+} &
            \DoReturn & \\
\multicolumn{5}{@{}l}{
\qquad
       \Forall{\Var{x^+}}{\IMu[\Interpretorn{\Var{O}}\: \Var{j}]}
            \Coherence[\Var{T^+}\: (\Var{f}\: (\ForgetOrn[\Var{x^+}]))\: (\Var{f^+} \Var{x^+})]} \\
\Coherence & 
            (\muOrnTimes{\Var{O}}{(\Inv[\Var{j}])}{\Var{T^+}}) &
            \Pair{\Var{x}}{\Var{xs}} &
            \Pair{\Var{x^+}}{\Var{xs^+}} &
            \DoReturn & \\
\multicolumn{5}{@{}l}{
\qquad
        \Var{x} \PropEqual \ForgetOrn[\Var{x^+}] \Times \Coherence[\Var{T^+}\: \Var{xs}\: \Var{xs^+}]} \\
\Coherence &
            \unit &
            \Void &
            \Void & 
            \DoReturn &
        \Unit
\end{array}
}
\]
}


\newcommand{\LookupType}{\Function{typeLookup}}

\begin{example}[Ornamenting \(\LeType\) to describe \(\Lookup\)]


In Section~\ref{sec:example-manual}, we have identified the ornaments
involved to transport the type of \(\_\Le\_\) to obtain the type of
\(\Lookup\). Let us spell them in full here. We need to ornament
\(\Nat\) into \(\List{A}\) and \(\Bool\) into \(\Maybe{A}\):
{
\[\Code{
\ListOrnDef \\
\\
\Ornament{\Maybe{}}
         {\Param{\Var{A}}{\Set}}
         {\Bool}{
\Emit{\Maybe{\Var{A}}}{}{\Just[\OrnInsert{\Var{a}}{\Var{A}}]}
                 \OrEmit{\Nothing}}
}
\]
}

From there, we give the functional ornament describing the
type of the \(\Lookup\) function:
{
\[
\Let{\LookupType}{\FunOrn[\LeType]}{
\Return{\LookupType}
       {\muOrnTo{\IdentityOrn{\Nat}}{\Void}{
         \muOrnTo{\OrnamentOf{\List{}}}{\Void}{
          \muOrnTimes{\OrnamentOf{\Maybe{}}}{\Void}
                     {\unit}}}}
}
\]
}
The user can verify that \(\interpretFunOrn{\LookupType}\)
gives us the type of the \(\Lookup\) function, up to multiplication by
\(\Unit\). Also, computing \(\Coherence[\LookupType\: (\_\Le\_)]\) gives
the expected result:
{
\[
\Code{
\LamAnn{\Var{f^+}}{\interpretFunOrn{\LookupType}}\\
\quad
 \Forall{\Var{n}}{\Nat}
   \Forall{\Var{xs}}{\List{A}} 
     \IsJust[(\Var{f^+}\: \Var{n}\:\Var{xs})] \PropEqual \Var{n} \Le \LengthList[\Var{xs}]
}
\]
}
Note that this equation is not \emph{specifying} the \(\Lookup\)
function: it is only establishing a computational relation between
\(\_\Le\_\) and a candidate lifting \(f^+\), for which \(\Lookup\) is
a valid choice. However, one could be interested in other functions
satisfying this coherence property and they would be handled by our
system just as well.

\end{example}

\newcommand{\AppendType}{\Function{type}\green{+\!+}}

\begin{example}[Ornamenting \(\PlusType\) to describe \(\_\Append\_\)]

The functional ornament of \(\PlusType\) makes only use of the
ornamentation of \(\Nat\) into \(\List{A}\):
{
\[
\Let{\AppendType}{\FunOrn[\PlusType]}{
\Return{\AppendType}
       {\muOrnTo{\OrnamentOf{\List{}}}{\Void}{
           \muOrnTo{\OrnamentOf{\List{}}}{\Void}{
             \muOrnTimes{\OrnamentOf{\List{}}}{\Void}
                        {\unit}}}}
}
\]
}
Again, computing \(\interpretFunOrn{\AppendType}\) indeed gives us the
type of \(\_\Append\_\) while \(\Coherence[\AppendType\:
  (\_\Plus\_)]\) correctly captures our requirement that list append
preserves the length of its arguments. As before, the list append
function is not the only valid lifting: one could for example consider
a function that reverts the first list and appends it to the second
one.

\end{example}


\subsection{Patches}


\Spacedcommand{\Patch}{\Function{Patch}}

By definition of a functional ornament, the lifting of a base function
\(\TypeAnn{f}{\interpretType{T}}\) is a function \(f^+\) of type
\(\interpretFunOrn{T^+}\) satisfying the coherence property
\(\Coherence[T^+\: f]\). To implement a lifting that is coherent, we
might ask the user to first implement the lifting \(f^+\) and then
prove it coherent. However, we find this process unsatisfactory: we
fail to harness the power of dependent types when implementing
\(f^+\), this weakness being then paid off by tedious proof
obligations. To overcome this limitation, we define the notion of
\(\Patch\) as the type of \emph{all} the functions that are coherent
by construction.

Note that we are looking for an equivalence here: we will define
patches so that they are in bijection with liftings satisfying a
coherence property, informally:
{
\begin{equation}\label{eqn:patch-coherence}
\Patch[T\: T^+\: f] \cong \SigmaTimes{\Var{f^+}}{\interpretFunOrn{T^+}}
                                     {\Coherence[T^+\: f\: \Var{f^+}]}
\end{equation}
}
In this paper, we constructively use this
bijection in the left to right direction: having implemented a patch
\(f^{++}\) of type \(\Patch[T\:\: T^+\: f]\), we will show, in the next
Section, how we can extract a lifting together with its coherence
proof.


Before giving the general construction of a \(\Patch\), let us first
work through the \(\_\Le\_\) example. After having functionally
ornamented \(\_\Le\_\) with \(\LookupType\), the lifting function
\(f^+\) and coherence property can be represented by the following
commuting diagram:
\begin{equation}\label{eqn:coherence-le}
\begin{tikzpicture}
\matrix (m) [matrix of math nodes
            , row sep=3em
            , column sep=5em
            , text height=1.5ex
            , text depth=0.25ex
            , ampersand replacement=\&]
{ 
   \Nat     \& \List{\Var{A}}   \& \Maybe{\Var{A}}    \\
   \Nat     \& \Nat             \& \Bool              \\
};
\path 
   (m-2-1) -- (m-2-2) node [midway] {\(\Times\)}
   (m-1-1) -- (m-1-2) node [midway] {\(\Times\)};
\draw[double,-]
   (m-1-1) -- node[left] { \(\Function{id}\) } (m-2-1);
\path[->] 
   (m-2-2) edge node[below] { \(\_\Le\_\) } (m-2-3)
   (m-1-2) edge node[above] { \(\green{f^+}\) } (m-1-3)
   (m-1-2) edge node[left] { \(\OrnForget{\CN{List}}\) } (m-2-2)
   (m-1-3) edge node[right] { \(\OrnForget{\CN{Maybe}}\) } (m-2-3);
\end{tikzpicture}
\end{equation}
In type theory, this is written as:
{
\[
\begin{array}{@{}l}
\SigmaTimes{\Var{f^+}}
           {\Nat \Times \List{A} \To \Maybe{A}}\\\:\:
           {\Forall{\Var{m}}{\Nat}
            \Forall{\Var{as}}{\List{A}}
            \Var{m} \Le \OrnForget{\CN{List}}\: \Var{as}
              \PropEqual 
            \OrnForget{\CN{Maybe}}\: (\Var{f^+}\: \Var{m}\: \Var{as})}
\end{array}
\]
}
Applying dependent choice, this is equivalent to:
{
\[
\cong
\begin{array}[t]{@{}l}
     \SigmaTimes{\Var{m}}{\Nat}
     \SigmaTimes{\Var{n}}{\Nat}
     \SigmaTimes{\Var{as}}{\List{A}}{\OrnForget{\CN{List}}\: \Var{as} \PropEqual \Var{n}} \To \\
\qquad \SigmaTimes{\Var{ma}}
                  {\Maybe{A}}
                  {\OrnForget{\CN{Maybe}}\: \Var{ma} \PropEqual \Var{m} \Le \Var{n}}
\end{array}
\]
}
Now, by definition of reornaments, we have that:
{
\begin{align*}
\SigmaTimes{\Var{as}}
           {\List{A}}
           {\OrnForget{\CN{List}}\: {\Var{as}} \PropEqual n} 
    & \cong \Vector{A}[n] 
    & \mbox{and} \\
\SigmaTimes{\Var{ma}}
           {\Maybe{A}}
           {\OrnForget{\CN{Maybe}}\: \Var{ma} \PropEqual b} 
    & \cong \IMaybe{A}[b]
\end{align*}
}
Applying these isomorphisms, we obtain the following type, which we
call the \(\Patch\) of the functional ornament \(\LookupType\):
{
\[
\cong
     \SigmaTimes{\Var{m}}{\Nat}
     \SigmaTimes{\Var{n}}{\Nat}
     \PiTo{\Var{vs}}{\Vector{A}[\Var{n}]}
     \IMaybe{A}[(\Var{m} \Le \Var{n})]
\]
}
Which is therefore equivalent to a pair of a lifting and its coherence
proof.

Intuitively, the \(\Patch\) construction consists in turning all the
vertical arrows of the commuting diagram (\ref{eqn:coherence-le}) into
the equivalent reornaments, or put in type theory: turning the pairs
of data and their algebraically defined constraint into equivalent
reornaments. The coherence property of reornaments tells us that
projecting the ornamented function down to its unornamented components
gives back the base function. By turning the projection functions into
inductive datatypes, we enforce the coherence property directly by
the index: we introduce a fresh index for the arguments (here,
introducing \(m\) and \(n\)) and index the return types by the result
of the unornamented function (here, indexing \(\IMaybe{A}\) by the
result \(m \Le n\)).




It is easy to generalise this construction to any functional
ornament. For clarity, we shall only write the proof for arity
one. The generalisation to multiple input and output arities is
straightforward but tediously verbose. So, from a base function
\(\Var{f}\), we start with its lifting and the associated coherence
property:
{
\[
\begin{array}{@{}l}
\SigmaTimes{\Var{f^+}}
           {\IMu[\Interpretorn{O_A}] \To \IMu[\Interpretorn{O_B}]} \\
\qquad
           \Forall{\Var{a^+}}{\IMu[\Interpretorn{O_A}]} 
             \OrnForget{O_B} (\Var{f^+}\: \Var{a^+})
               \PropEqual
             \Var{f} (\OrnForget{O_A}\: \Var{a^+})
\end{array}
\]
}
Applying dependent choice, we obtain the following equivalent type:
{
\[
\cong 
 \SigmaTimes{\Var{a}}{\IMu[A]}
 \SigmaTimes{\Var{a^+}}{\IMu[\Interpretorn{O_A}]}
 \OrnForget{O_A}\: \Var{a^+} \PropEqual \Var{a} 
\To
 \SigmaTimes{\Var{b^+}}{\IMu[\Interpretorn{O_B}]}
 \OrnForget{O_B} \Var{b^+} 
    \PropEqual
 \Var{f} \Var{a}
\]
}
Then, we can simply use the characterisation of a reornament to turn
every pair
\(\SigmaTimes{\Var{x^+}}{\IMu[\Interpretorn{O_X}]}{\OrnForget{O_X}\:
  \Var{x^+} \PropEqual t}\) into the equivalent inductive type
\(\IMu[\OrnAlgOrn{X}{O_X}\: t]\):
{
\[
\begin{array}{@{}l@{\:}l}
\cong &
 \SigmaTimes{\Var{a}}{\IMu[A]}
            {\IMu[\OrnAlgOrn{A}{O_{A}}\: \Var{a}]} \To
   \IMu[\OrnAlgOrn{B}{O_B}] (\Var{f}\: \Var{a})
\end{array}
\]
}

To build this type generically, we simply proceed by induction over
the functional ornament. Upon an argument (\ie a \(\muOrnTo{O}{}{}
\)), we introduce a fresh index and the reornament of \(O\). Upon a
result (\ie a \(\muOrnTimes{O}{}{}\)), we ask for a reornament of
\(O\) indexed by the result of the base function.
{
\[
\Let{
\Patch &
         \PiTel{\Var{T}}{\TypeUniv} &
         \PiTel{\Var{T^+}}{\FunOrn[\Var{T}]} &
         \PiTel{\Var{f}}{\interpretType{\Var{T}}}}
    {\Set}{
\Return{\Patch &
            (\muTo{\Var{D}}{(\Var{re}\:\Var{j})}{\Var{T}}) &
            (\muOrnTo{\Var{O}}{(\Inv[\Var{j}])}{\Var{T^+}}) &
            \Var{f}}
       {}
\multicolumn{6}{l}{
\qquad\PiTo{\Var{x}}{\IMu[\Var{D}\: (\Var{re}\: \Var{j})]}
         \IMu[\OrnAlgOrn{\Var{D}}{\Var{O}}\: \Pair{\Var{j}}{\Var{x}}] \To
         \Patch[\Var{T}\: \Var{T^+}\: (\Var{f}\: \Var{x})]} \\
\Return{\Patch &
            (\muTimes{\Var{D}}{(\Var{re}\:\Var{j})}{\Var{T}}) &
            (\muOrnTimes{\Var{O}}{(\Inv[\Var{j}])}{\Var{T^+}}) &
            \Pair{\Var{x}}{\Var{xs}}}
       {}
\multicolumn{6}{l}{
\qquad\IMu[\OrnAlgOrn{\Var{D}}{\Var{O}}\: \Pair{\Var{j}}{\Var{x}}] \Times \Patch[\Var{T}\: \Var{T^+}\: \Var{xs}]}\\
\Return{\Patch & 
            \unit &
            \unit &
            \Void}
       {\Unit}
}\]
}


\begin{example}[\(\Patch\) of \(\LookupType\)]


The type of the coherent liftings of \(\_\Le\_\) by \(\LookupType\),
as defined by the \(\Patch\) of \(\_\Le\_\) by \(\LookupType\),
computes to:
{
\[
  \PiTel{\Var{m}}{\Nat} 
  \PiTo{\Var{m^+}}{\IMu[\OrnAlgOrn{\Nat}{\IdentityOrn{\Nat}}\: \Var{m}]} 
    \PiTel{\Var{n}}{\Nat}
    \PiTo{\Var{vs}}{\IMu[\OrnAlgOrn{\Nat}{\List{A}}\: \Var{n}]} 
      \IMu[\OrnAlgOrn{\Bool}{\Maybe{A}}\: (\Var{m} \Le \Var{n})] \Times \Unit
\]
}

Note that \(\IMu[\OrnAlgOrn{\Nat}{\IdentityOrn{\Nat}}\: n]\) is
isomorphic to \(\Unit\): all the content of the datatype has been forced
-- the recursive structure of the datatype is entirely determined by its
index -- and detagged -- the choice of constructors is entirely
determined by its index, leaving no actual data in it. Hence, we
discard this argument as computationally uninteresting. On the other
hand, \(\OrnAlgOrn{\Nat}{\List{A}}\) and
\(\OrnAlgOrn{\Bool}{\Maybe{A}}\) are, respectively, the
previously introduced \(\Vector{A}\) and \(\IMaybe{A}\)
types.

\end{example}

\begin{example}[\(\Patch\) of \(\PlusType\)]

Similarly, the \(\Patch\) of \(\_\Plus\_\) by \(\PlusType\) computes
to:
{
\[
\PiTel{\Var{m}}{\Nat} \PiTo{\Var{xs}}{\OrnAlgOrn{\Nat}{\List{A}}\: \Var{m}}
  \PiTel{\Var{n}}{\Nat} \PiTo{\Var{ys}}{\OrnAlgOrn{\Nat}{\List{A}}\: \Var{m}}
        \OrnAlgOrn{\Nat}{\List{A}}\: (\Var{m} \Plus \Var{n}) \Times \Unit
\]
}
Which is the type of the vector append function.

\end{example}



\subsection{Patching and coherence}



\Spacedcommand{\patch}{\Function{patch}}
\Spacedcommand{\coherence}{\Function{coherence}}

At this stage, we can implement the \(\Ilookup\) function exactly as
we did in Section~\ref{sec:example-manual}. From there, we now want to
obtain the \(\Lookup\) function and its coherence certificate. More
generally, having implemented a function satisfying the \(\Patch\)
type, we want to extract the lifting and its coherence proof. 

Perhaps not surprisingly, we obtain this construction by looking at
the isomorphism~(\ref{eqn:patch-coherence}) of the previous Section
through our constructive glasses: indeed, as the \(\Patch\) type is
isomorphic to the set of liftings satisfying the coherence property,
we effectively get a function taking every \(\Patch\) to a lifting and
its coherence proof. More precisely, we obtain the lifting by
generalising the reornament-induced \(\OrnForget{*}\) maps to
functional ornaments while we obtain the coherence proof by
generalising the reornament-induced \(\OAAO\) theorem to functional
ornaments.

We call \emph{patching} the action of projecting the coherent lifting
from a \(\Patch\) function. Again, it is defined by mere induction
over the functional ornament. When ornamented arguments are introduced
(\ie with \(\muOrnTo{O}{}{}\)), we simply patch the body of the
function. This is possible because from
\(\TypeAnn{x^+}{\IMu[\Interpretorn{O_D}]}\), we can forget the
ornament to compute \(f\: (\ForgetOrn[x^+])\) and we can also make the
reornament to compute \(f^{++}\: \_\: (\MakeAlgOrn[x^+])\). When an
ornamented result is to be returned, we simply forget the
reornamentation computed by the coherent lifting:
{
\[
\Let{\patch &
         \PiTel{\Var{T^+}}{\FunOrn[\Var{T}]} &
         \PiTel{\Var{f}}{\interpretType{\Var{T}}} &
         \PiTel{\Var{p}}{\Patch[\Var{T}\: \Var{T^+}\: \Var{f}]}}
    {\interpretFunOrn{\Var{T^+}}}{
\Return{\patch &
            (\muOrnTo{\Var{O}}{(\Inv[\Var{j}])}{\Var{T^+}}) &
            \Var{f} &
            \Var{f^{++}}}{}
\multicolumn{6}{l}
       {\qquad\Lam{\Var{x^+}}{\begin{array}[t]{@{}l@{\:}l}
                       \patch & (\Var{f}\: (\ForgetOrn[\Var{x^+}])) \\
                              & (\Var{f^{++}}\: (\ForgetOrn[\Var{x^+}])\: (\MakeAlgOrn[\Var{x^+}]))
                       \end{array}}} \\
\Return{\patch &
            (\muOrnTimes{\Var{O}}{(\Inv[\Var{j}])}{\Var{T^+}}) &
            \Pair{\Var{x}}{\Var{xs}} &
            \Pair{\Var{x^{++}}}{\Var{xs^{++}}}}{}
\multicolumn{6}{l}{
       \qquad\Pair{\ForgetOrn[\Var{x^{++}}]}{\patch[\Var{T^+}\: \Var{xs}\: \Var{xs^{++}}]}} \\
\Return{\patch &
            \unit &
            \Void &
            \Void}
       {\Void}
}
\]
}

Extracting the coherence proof follows a similar pattern. We introduce
arguments as we go, just as we did with \(\patch\). When we reach a
result, we have to prove the coherence of the result returned by the
patched function: this is a straightforward application of the
\(\OAAO\) theorem:
{
\[
\Let{\coherence &
         \PiTel{\Var{T^+}}{\FunOrn[\Var{T}]} &
         \PiTel{\Var{f}}{\interpretType{\Var{T}}} &
         \PiTel{\Var{p}}{\Patch[\Var{T}\: \Var{T^+}\: \Var{f}]}}
    {\Coherence[\Var{T^+}\: \Var{f}\: (\patch[\Var{T^+}\: \Var{f}\: \Var{p}])]}{
\Return{\coherence &
            (\muOrnTo{\Var{O}}{(\Inv[\Var{j}])}{\Var{T^+}}) &
            \Var{f} &
            \Var{p}}{}
\multicolumn{6}{l}{
       \qquad\Lam{\Var{x^+}}{\begin{array}[t]{@{}l@{\:}l}
                       \coherence & \Var{T^+} \\
                                  & (\Var{f}\: (\ForgetOrn[\Var{x^+}])) \\
                                  & (\Var{p}\: (\ForgetOrn[\Var{x^+}])\: (\MakeAlgOrn[\Var{x^+}]))
                       \end{array}}}\\
\Return{\coherence &
            (\muOrnTimes{\Var{O}}{(\Inv[\Var{j}])}{\Var{T^+}}) &
            \Pair{\Var{x}}{\Var{xs}} &
            \Pair{\Var{x^+}}{\Var{p}}}{}
\multicolumn{6}{l}{
  \qquad\Pair{\OAAO[\Var{x^+}]}{\coherence[\Var{T^+}\: \Var{xs}\: \Var{p}]}}\\
\Return{\coherence &
            \unit &
            \Void &
            \Void}{\Void}
}\]}


\begin{example}[Obtaining \(\Lookup\) and its coherence certificate, for free]


This last step is a mere application of the \(\patch\) and
\(\coherence\) functions. Hence, we define \(\Lookup\) as follows:
{
\[
\Let{\Lookup}{\interpretFunOrn{\LookupType}}{
\Return{\Lookup}{\patch[\LookupType\: (\_\Le\_)\: \Ilookup]}
}
\]
}
And we get its coherence proof, here spelled in full:
{
\[
\Let{\CohLookup & 
         \PiTel{\Var{n}}{\Nat} &
         \PiTel{\Var{xs}}{\List{\Var{A}}}}
    {\OrnForget{\CN{Maybe}}\: (\Fst (\Lookup[\Var{n}\:\Var{xs}])) 
      \PropEqual 
      \Fst (\Var{n} \Le \OrnForget{\CN{List}}\: \Var{xs})}{
\Return{\CohLookup & \Var{n} & \Var{xs}}{\coherence[\LookupType\: (\_\Le\_)\: \Ilookup]\: \Var{n}\: \Var{xs}}}
\]}
\end{example}

\Spacedcommand{\VectorAppend}{\Function{vappend}}
\Spacedcommand{\CohAppend}{\Function{coh}\green{+\!+}}

\begin{example}[Obtaining \(\_\Append\_\) and its coherence certificate, for free]

Assuming that we have implemented the coherent lifting
\(\VectorAppend\), we obtain concatenation of lists and its coherence
proof by simply running our generic machinery:
{
\[
\Code{
\Let{\Append}{\interpretFunOrn{\AppendType}}{
\Return{\Append}{\patch[\AppendType\: (\_\Plus\_)\: \VectorAppend]}
} \\
\\
\Let{\CohAppend & 
         \PiTel{\Var{xs}}{\List{\Var{A}}} &
         \PiTel{\Var{ys}}{\List{\Var{A}}}}
    {\OrnForget{\CN{List}}\: (\Fst (\Var{xs} \Append \Var{ys})) 
      \PropEqual 
      \Fst ((\OrnForget{\CN{List}}\: \Var{xs}) \Plus (\OrnForget{\CN{List}}\: \Var{ys}))}{
\Return{\CohAppend & \Var{xs} & \Var{ys}}{
  \coherence[\AppendType\: (\_\Plus\_)\: \VectorAppend]\: \Var{xs}\: \Var{ys}}}
}
\]
}

\end{example}


Looking back at the manual construction in
Section~\ref{sec:example-manual}, we can measure the progress we have
made: while we had to duplicate entirely the type signature of
\(\Lookup\) and its coherence proof, we can now write down a
functional ornament and these are generated for us. This is not just
convenient: by giving a functional ornament, we establish a strong
connection between two functions. By pinning down this connection with
the universe of functional ornaments, we turn this knowledge into an
effective object that can be manipulated and reasoned about within the
type theory.

We make use of this concrete object when we construct the \(\Patch\)
induced by a functional ornament: this is again a construction that is
generic now, while we had to tediously (and perhaps painfully)
construct it in Section~\ref{sec:example-manual}. Similarly, we get
patching and extraction of the coherence proof for free now, while we
had to manually fiddle with several projection and injection
functions.

We presented the \(\Patch\) as the type of the liftings coherent by
construction. As we have seen, its construction and further projection
down to a lifting is now entirely automated, hence effortless. This is
a significant step forward: we could either implement \(\Lookup\) and
then prove it coherent, or we could go through the trouble of manually
defining carefully indexed types and write a function correct by
construction. We have now made this second alternative just as
accessible as the first one. And, from a programming
perspective, the second approach is much more appealing. In a word, we
have made an appealing technique extremely cheap!

Finally, we shall reiterate that none of the above constructions
involve extending the type theory: for a theory with a universe of
datatypes, the theory of functional ornaments can be entirely
internalised as a few generic programs and inductive types. For a
system lacking a universe of datatypes, this technology would need to
be provided at the meta-level. However, the fact that our
constructions type-check in our system suggests that adding
these constructions at the meta-level is consistent with a
pre-existing meta-theory.


\section{Lazy programmers, clever constructors}
\label{sec:clever-constructors}


\begin{figure}[tb]
{\small
\[
\begin{array}{l@{\qquad\qquad}r}
\LeExplicitDef & 
\Let{\Ilookup &
         \PiTel{\Var{m}}{\Nat} & 
         \PiTel{\Var{vs}}{\Vector{\Var{A}}[\Var{n}]}}
    {\IMaybe{\Var{A}}[(\Var{m} \Le \Var{n})]}{
\By{\Ilookup & \Var{m} & \Var{vs}}{\VectorCase[\Var{vs}]}{
  \Return{\quad \Ilookup & \Var{m} & \VNil}{\Nothing}
  \By{\quad\Ilookup & \Var{m} & (\VCons[\Var{a}\: \Var{vs}])}{\NatElim[\Var{m}]}{
    \Return{\quad\quad \Ilookup & \Zero & (\VCons[\Var{a}\: \Var{vs}])}{\Just[\Var{a}]}
    \Return{\quad\quad \Ilookup & (\Suc[\Var{m}]) & (\VCons[\Var{a}\: \Var{vs}])}{\Ilookup[\Var{m}\: \Var{vs}]}
}}}
\end{array}
\]}
\caption{Implementations of \(\_\Le\_\) and \(\Ilookup\)}
\label{fig:le-vs-lookup}
\end{figure}

In our journey from \(\_\Le\_\) to \(\Lookup\), we had to implement
the \(\Ilookup\) function. It is instructive to put \(\_\Le\_\) and
\(\Ilookup\) side-by-side (Fig.~\ref{fig:le-vs-lookup}).  First, both
functions follow the same recursion pattern: case analysis over
\(n\)/\(vs\) followed by induction over \(m\). Second, the returned
constructors are related through the \(\Maybe{}\) ornament: knowing
that we have returned \(\True\) or \(\False\) when implementing
\(\_\Le\_\), we can deduce which of \(\Just\) or \(\Nothing\) will be
used in \(\Ilookup\). Interestingly, the only unknown, hence the only
necessary input from the user, is the \(a\) in the \(\Just\) case: it
is precisely the information that has been introduced by the
\(\Maybe{}\) ornament.


In this Section, we are going to leverage our knowledge of the
definition of the base function -- such as \(\_\Le\_\) -- to guide the
implementation of the coherent lifting -- such as \(\Ilookup\):
instead of re-implementing \(\Ilookup\) by duplicating most of the
code of \(\_\Le\_\), the user indicates \emph{what to duplicate} and
only provides \emph{strictly necessary} inputs. We are primarily
interested in transporting two forms of structure:
\begin{description}
\item[Recursion pattern:] if the base function is a fold
  \(\Fold{\alpha}\) and the user provides us with a
  \emph{coherent algebra} \(\hat{\beta}\) of
  \(\alpha\), we automatically construct the coherent lifting
  \(\Fold{\hat{\beta}}\) of \(\Fold{\alpha}\) ;
\item[Returned constructor:] if the base function returns a
  constructor \(C\) and the user provides us with a
  \emph{coherent extension} \(\hat{C}\) of \(C\), we automatically
  construct the coherent lifting of \(C\)
\end{description}
We shall formalise what we understand by being a coherent algebra and
a coherent extension below. The key idea is to identify the strictly
necessary inputs from the user, helped in that by the ornaments. It is
then straightforward to, automatically and generically, build the
lifted folds and values.


\subsection{Transporting recursion patterns}

\Spacedcommand{\LiftFold}{\Function{lift-fold}}
\Spacedcommand{\LiftInd}{\Function{lift-ind}}
\Spacedcommand{\LiftCase}{\Function{lift-case}}

\Spacedcommand{\IsSuc}{\Function{isSuc}}
\Spacedcommand{\IsSucAlg}{\green{\alpha_{\IsSuc}}}
\Spacedcommand{\Head}{\Function{hd}}
\Spacedcommand{\HeadAlg}{\green{\alpha_{\Head}}}
\Spacedcommand{\Ihead}{\Function{ihd}}
\Spacedcommand{\IheadAlg}{\green{\alpha_{\Ihead}}}

\newcommand{\LengthListNat}[1]{\OrnForgetNat{\CN{List}}}

\Spacedcommand{\LiftConstructor}{\Function{lift-constructor}}

\begin{figure}[tb]
{
\[
\Code{
\mbox{(a) Request lifting of algebra (user input):} \\
\\
\Let{\Ihead & \PiTel{\Var{vs}}{\Vector{\Var{A}}[\Var{n}]}}
    {\IMaybe{\Var{A}}[\IsSuc[\Var{n}]]}{
\LiftBy{\Ihead & }
        {\LiftFold}{\Hole}}
\\
\\
\mbox{(b) Result of lifting the algebra (system output):} \\
\\
\Code{
\Let{\Ihead & \PiTel{\Var{vs}}{\Vector{\Var{A}}[\Var{n}]}}
    {\IMaybe{\Var{A}}[(\IsSuc[\Var{n}])]}{
\LiftBy{\Ihead & }
        {\LiftFold \quad\where}{}} \\
\begin{array}{@{\qquad}l}
   \Let{\IheadAlg & \PiTel{\Var{vs}}{\InterpretIDesc{\IDescOf{\Vector{}}}[(\Lam{\Var{n'}}\IMaybe{\Var{A}}[(\IsSuc[\Var{n'}])])]\: \Var{n}}}
       {\IMaybe{\Var{A}}[(\IsSuc[\Var{n}])]}{
   \HoleCommand{\IheadAlg & \Tag{\CN{nil}}}
   \HoleCommand{\IheadAlg & (\Tag{\CN{cons}}\: \Var{a}\: \Var{xs})}
\end{array}}}
\\
\\
\mbox{(c) Request lifting of constructors (user input):} \\
\\
\Code{
\Let{\Ihead & \PiTel{\Var{vs}}{\Vector{\Var{A}}[\Var{n}]}}
    {\IMaybe{\Var{A}}[(\IsSuc[\Var{n}])]}{
\LiftBy{\Ihead & }
        {\LiftFold \quad\where}{}}\\
\begin{array}{@{\qquad}l}
   \Let{\IheadAlg & \PiTel{\Var{vs}}{\InterpretIDesc{\IDescOf{\Vector{}}}[(\Lam{\Var{n'}}\IMaybe{\Var{A}}[(\IsSuc[\Var{n'}])])]\: \Var{n}}}
       {\IMaybe{\Var{A}}[(\IsSuc[\Var{n}])]}{
   \IheadAlg & \Tag{\CN{nil}} & \DoLiftReturn & \Hole \hfill \\
   \IheadAlg & (\Tag{\CN{cons}}\: \Var{a}\: \Var{xs}) & \DoLiftReturn & \Hole \hfill
\end{array}
}}
\\
\\
\mbox{(d) Result of lifting constructors (system output)} \\
\\
\Code{
\Let{\Ihead & \PiTel{\Var{vs}}{\Vector{\Var{A}}[\Var{n}]}}
    {\IMaybe{\Var{A}}[(\IsSuc[\Var{n}])]}{
\LiftBy{\Ihead & }
        {\LiftFold \quad\where}{}} \\
\begin{array}{@{\qquad}l}
   \Let{\IheadAlg & \PiTel{\Var{vs}}{\InterpretIDesc{\IDescOf{\Vector{}}}[(\Lam{\Var{n'}}\IMaybe{\Var{A}}[(\IsSuc[\Var{n'}])])]\: \Var{n}}}
       {\IMaybe{\Var{A}}[(\IsSuc[\Var{n}])]}{
   \LiftReturn{\IheadAlg & \Tag{\CN{nil}}}{\Nothing\: \HoleAnn{\Unit}}{\HoleAnn{\Unit}}{}
   \LiftReturn{\IheadAlg & (\Tag{\CN{cons}}\: \Var{a}\: \Var{xs})}{\Just[\HoleAnn{\Var{A}}]}{\HoleAnn{\Unit}}{}
\end{array}
}}
\\
\\
\mbox{(e) Type-checked term (automatically generated from (d)):} \\
\\
\Code{
\Let{\Ihead & \PiTel{\Var{vs}}{\Vector{\Var{A}}[\Var{n}]}}
    {\IMaybe{\Var{A}}[(\IsSuc[\Var{n}])]}{
\Return{\Ihead & \Var{vs}}{\LiftFold[\IsSucAlg\: \IheadAlg]  \quad\where}}\\
\begin{array}{@{\qquad}l}
   \Let{\IheadAlg & \PiTel{\Var{vs}}{\InterpretIDesc{\IDescOf{\Vector{}}}[(\Lam{\Var{n'}}\IMaybe{\Var{A}}[(\IsSuc[\Var{n'}])])]\: \Var{n}}}
       {\IMaybe{\Var{A}}[(\IsSuc[\Var{n}])]}{
   \Return{\IheadAlg & \Tag{\CN{nil}}}{\LiftConstructor[\Tag{\CN{nil}}\: \HoleAnn{\Unit}\: \HoleAnn{\Unit}\: \Void]}
   \Return{\IheadAlg & (\Tag{\CN{cons}}\: \Var{a}\: \Var{xs})}{\LiftConstructor[(\Tag{\CN{suc}}\: \Var{n})\: \HoleAnn{\Var{A}}\: \HoleAnn{\Unit}\: \Void]}
\end{array}
}}
}
\]
}

\caption{Guided implementation of \(\Ihead\)}
\label{fig:ihead-implem}

\end{figure}


When transporting a function, we are very unlikely to change the
recursion pattern of the base function. Indeed, the very reason why we
\emph{can} do this transportation is that the lifting uses exactly the
same structure to compute its results. Hence, in the majority of the
cases, we could just ask the computer to use the induction principle
induced by the base one: the only task left to the user will be to
give the algebra. For clarity of exposition, we restrict ourselves to
transporting folds. However, the treatment of induction is essentially
the same, as hinted by the fact that induction can be reduced to
folds~\citep{fumex:fibrational-induction}.

To understand how we transport the recursion pattern, let us look
again at the coherence property of liftings, but this time
specialising to functions that are folds:
\[
\begin{tikzpicture}
\matrix (m) [matrix of math nodes
            , row sep=3em
            , column sep=5em
            , text height=1.5ex
            , text depth=0.25ex
            , ampersand replacement=\&]
{ 
   \IMu[\Interpretorn{O_D}] \& \interpretFunOrn{T^+}     \\
   \IMu[D]                  \& \interpretType{T}         \\
};
\path[->] 
   (m-1-1) edge node[left] { \(\OrnForget{O_D}\) } (m-2-1)
   (m-1-2) edge node[right] { \(\OrnForget{T^+}\) } (m-2-2)
   (m-2-1) edge node[below] { \(\Fold{\alpha}\) } (m-2-2)
   (m-1-1) edge node[above] { \(\Fold{\beta}\) } (m-1-2);
\end{tikzpicture}
\]
By the fold-fusion theorem~\citep{bird:algebra-of-programming}, it is
sufficient (but not necessary) to work on the algebras, where we have
the following diagram:
%
\[
\begin{tikzpicture}
\matrix (m) [matrix of math nodes
            , row sep=3em
            , column sep=5em
            , text height=1.5ex
            , text depth=0.25ex
            , ampersand replacement=\&]
{ 
   \InterpretIDesc{\Interpretorn{O_D}}[\interpretFunOrn{T^+}] 
       \& 
       \& \interpretFunOrn{T^+}     \\
   \InterpretIDesc{\Interpretorn{O_D}}[\interpretType{T}]
       \& \InterpretIDesc{D}[\interpretType{T}]
       \& \interpretType{T}                      \\
};
\path[->] 
   (m-1-1) edge node[right] { \(\InterpretIDesc{O_D}[\OrnForget{T^+}]\) } (m-2-1)
   (m-2-1) edge node[below] { \(\OrnForgetNat{O_D}\) } (m-2-2)
   (m-1-3) edge node[left] { \(\OrnForget{T^+}\) } (m-2-3)
   (m-2-2) edge node[below] { \(\alpha\) } (m-2-3)
   (m-1-1) edge node[above] { \(\beta\) } (m-1-3);
\end{tikzpicture}
\]


Now, we would like to find an algebra \(\hat{\beta}\) such that its
fold gives us a function of the \(\Patch\) type.

To illustrate this approach, we work through a concrete example: we
derive \(\TypeAnn{\Head}{\List{A} \To \Maybe{A}}\) from
\(\TypeAnn{\IsSuc}{\Nat \To \Bool}\) by transporting the algebra. For
the sake of argument, we artificially define \(\IsSuc\) by a fold:
{
\[
\Code{
\Let{\IsSuc & \PiTel{\Var{n}}{\Nat}}
    {\Bool}{
\Return{\IsSuc & \Var{n}}{\Fold{\IsSucAlg}\: \Var{n} \quad\where}}\\
\begin{array}{@{\qquad}l}
   \Let{\IsSucAlg & \PiTel{\Var{xs}}{\InterpretIDesc{\IDescOf{\Nat}}[\Bool]}}
       {\Bool}{
   \Return{\IsSucAlg & \Tag{\CN{0}}}{\False}
   \Return{\IsSucAlg & (\Tag{\CN{suc}}\: \Var{xs})}{\True}}
\end{array}}
\]
}

Our objective is thus to define the algebra for \(\Head\), which has
the following type:
{
\[
\TypeAnn{\HeadAlg}{\InterpretIDesc{\IDescOf{\List{}}}[\Maybe{A}] \To \Maybe{A}}
\]
}
such that its fold is coherent. By the fold-fusion
theorem~\citep{bird:algebra-of-programming}, it is sufficient (but not
necessary) for \(\HeadAlg\) to satisfy the following condition:
{
\[
\Code{
\Forall{\Var{ms}}{\InterpretIDesc{\IDescOf{\List{}}}[\Maybe{A}]} \\
\quad\:\:     \IsJust[(\HeadAlg\: \Var{ms})]
             \PropEqual 
           \IsSucAlg[(\LengthListNat{A}{(\InterpretIDesc{\IDescOf{\List{}}}[\IsJust]\: \Var{ms})})]
}
\]
}
Following the same methodology we applied to define the \(\Patch\)
type, we can massage the type of \(\HeadAlg\) and its coherence
condition to obtain an equivalent definition enforcing the coherence
by indexing. In this case, the natural candidate is:
{
\[
\TypeAnn{\IheadAlg}{
          {\InterpretIDesc{\IDescOf{\Vector{}}}
                          [(\Lam{\Var{n'}}\IMaybe{A}[(\IsSuc[\Var{n'}])])]\: \Var{n}} 
          \To
               \IMaybe{A}[(\IsSuc[\Var{n}])]}
\]
}


This construction generalises to any functional ornament. That is, from
an algebra
{
\[
\TypeAnn{\Var{\alpha}}
        {\PiTo{\Var{i}}{I} 
               \InterpretIDesc{D\: \Var{i}}[(\Lam{\_}{\interpretType{T}})] \To 
               \interpretType{T}}
\]
}
together with an ornament \(\TypeAnn{O_D}{\orn[re\: D]}\)
and a functional ornament \(\TypeAnn{T^+}{\FunOrn[T]}\), the type
of coherent algebras for \(\alpha\) is:
{
\[\Code{
\TypeAnn{\Var{\hat{\beta}}}{}
\PiTel{\Var{j}}{J}
\PiTo{\Var{t}}{\IMu[D\: (re\: \Var{j})]} \\
\qquad  \InterpretIDesc{\OrnAlgOrn{D}{O}\: \Pair{\Var{j}}{\Var{t}}}
                      [(\Lam{\Pair{\Var{j}}{\Var{t}}}
                            {\Patch[T\: (\Fold{\alpha}\: \Var{t})\: T^+]})] \To \\
\qquad\quad
         \Patch[T\: (\Fold{\alpha}\: \Var{t})\: T^+]
}\]
}
It can formally be proved that algebras of this type capture exactly
the algebras satisfying the coherence condition. Constructively, we
get that such a coherent algebra induces a coherent lifting, by a mere
fold of the coherent algebra:
{
\[
\begin{array}{@{}l@{\:}l}
\LiftFold & 
       \PiTel{\Var{\alpha}}
             {\PiTo{\Var{i}}{\Var{I}} 
               \InterpretIDesc{\Var{D}\: \Var{i}}[(\Lam{\_}{\interpretType{\Var{T}}})] \To 
               \interpretType{\Var{T}}} \\
          & 
       \PiTel{\Var{\hat{\beta}}}
             {\begin{array}[t]{l}
                 \PiTel{\Var{j}}{\Var{J}}
                 \PiTo{\Var{t}}{\IMu[\Var{D}\: (\Var{re}\: \Var{j})]} \\
               \quad  \InterpretIDesc{\OrnAlgOrn{\Var{D}}{\Var{O}}\: \Pair{\Var{j}}{\Var{t}}}
                                [(\Lam{\Pair{\Var{j}}{\Var{t}}}
                                      {\Patch[\Var{T}\: (\Fold{\Var{\alpha}}\: \Var{t})\: \Var{T^+}]})] \To \\
               \quad  \Patch[\Var{T}\: (\Fold{\Var{\alpha}}\: \Var{t})\: \Var{T^+}]}
               \end{array}
               \\
          & 
 \::\: \Patch[(\muTo{\Var{D}}{(\Var{re}\: \Var{j})}{\Var{T}})\: 
            \Fold{\Var{\alpha}}\: 
            (\muOrnTo{\Var{O}}{\Var{j}}{\Var{T^+}})] \\
\LiftFold & \Var{\alpha}\: \Var{\hat{\beta}} \DoReturn 
    \Lam{\Var{x}} 
    \Lam{\Var{x^{++}}}
      \Fold{\Var{\hat{\beta}}}\: \Var{x^{++}}
\end{array}
\]
}

Generalising this idea, we similarly lift induction:
{
\[
\begin{array}{@{}l@{\:}l}
\LiftInd & 
       \PiTel{\Var{\alpha}}
             {\begin{array}[t]{@{}l}
               \PiTel{\Var{i}}{\Var{I}}
               \PiTo{\Var{xs}}{\InterpretIDesc{\Var{D}\: \Var{i}}[\IMu[\Var{D}]]} \\
               \IAll[\Var{D}\: (\Lam{\_}{\interpretType{\Var{T}}})\: \Var{xs}] \To
               \interpretType{\Var{T}}}
               \end{array}
               \\
          & 
       \begin{array}[t]{@{}l@{\:}l}
       \PiTel{\Var{\hat{\beta}}}{&
              \PiTel{\Var{j}}{\Var{J}}
              \PiTel{\Var{t}}{\IMu[\Var{D}\: (\Var{re}\: \Var{j})]}
              \PiTo{\Var{xs}}{\InterpretIDesc{\OrnAlgOrn{\Var{D}}
                                                        {\Var{O}}\: 
                                                        \Pair{\Var{j}}{\Var{t}}}
                                             [\IMu[\OrnAlgOrn{\Var{D}}{\Var{O}}]]} \\
            & \IAll[(\OrnAlgOrn{\Var{D}}{\Var{O}}\:  
                    (\Lam{\Pair{\Var{j}}{\Var{t}}}{\Patch[\Var{T}\: ((\Iinduction[\_\:\_\:\Var{\alpha}])\: \Var{t})\: \Var{T^+}]})\:
                    \Var{xs}] \To \\
            & \Patch[\Var{T}\: ((\Iinduction[\_\:\_\:\Var{\alpha}])\: \Var{t})\: \Var{T^+}]}
       \end{array} \\
          & 
 \::\: \Patch[(\muTo{\Var{D}}{(\Var{re}\: \Var{j})}{\Var{T}})\: 
            (\Iinduction[\_\:\_\:\Var{\alpha}])\: 
            (\muOrnTo{\Var{O}}{\Var{j}}{\Var{T^+}})] \\
\LiftInd & \alpha\: \hat{\beta} \DoReturn 
    \Lam{\Var{x}} 
    \Lam{\Var{x^{++}}}
      \Iinduction[\_\:\_\:\hat{\beta}]\: \Var{x^{++}}
\end{array}
\]
}

Lifting case analysis is now simple, as case analysis is a specialisation of
induction where the induction hypotheses are stripped
out~\citep{mcbride:constructions-on-constructors}:

{
\[
\begin{array}{@{}l@{\:}l}
\LiftCase & 
       \PiTel{\Var{\alpha}}
             {\PiTel{\Var{i}}{I} 
               \PiTo{\Var{xs}}{\InterpretIDesc{\Var{D}\: \Var{i}}[\IMu[\Var{D}]]}
               \interpretType{\Var{T}}} \\
          & 
       \PiTel{\Var{\hat{\beta}}}{
              \begin{array}[t]{@{}l}
              \PiTel{\Var{j}}{\Var{J}}
              \PiTel{\Var{t}}{\IMu[\Var{D}\: (\Var{re}\: \Var{j})]}\\
\quad         \PiTo{\Var{xs}}{\InterpretIDesc{\OrnAlgOrn{\Var{D}}
                                                        {\Var{O}}\: 
                                                        \Pair{\Var{j}}{\Var{t}}}
                                             [\IMu[\OrnAlgOrn{\Var{D}}{\Var{O}}]]} \\
\quad         \Patch[\Var{T}\: ((\Iinduction[\_\:\_\:(\Lam{\Var{xs}\: \_}\Var{\alpha}\:\Var{xs})])\: \Var{t})\: \Var{T^+}]}
              \end{array}\\
          & 
 \::\: \begin{array}[t]{@{}l@{\:}l}
        \Patch & (\muTo{\Var{D}}{(\Var{re}\: \Var{j})}{\Var{T}}) \\
               & (\Iinduction[\_\:\_\:(\Lam{\Var{xs}\: \_}\Var{\alpha}\:\Var{xs})]) \\
               & (\muOrnTo{\Var{O}}{\Var{j}}{\Var{T^+}})
       \end{array} \\
\LiftCase & \alpha\: \hat{\beta} \DoReturn 
  \LiftInd[(\Lam{\Var{xs}\: \_}\Var{\alpha}\:\Var{xs})\: (\Lam{\Var{xs}\: \_}\Var{\hat{\beta}}\: \Var{xs})]
\end{array}
\]
}



\begin{example}[Transporting the recursion pattern of \(\IsSuc\)]

We can now apply our generic machinery to transport \(\IsSuc\) to
\(\Head\): in a high-level notation, we would write the command of
Fig.~\ref{fig:ihead-implem}(a). To this command, an interactive system
would respond by automatically generating the algebra, as shown in
Fig.~\ref{fig:ihead-implem}(b). In the low-level type theory, this
would elaborate to the following term:
{
\[
\Code{
\Let{\Ihead & \PiTel{\Var{vs}}{\Vector{\Var{A}}[\Var{n}]}}
    {\IMaybe{\Var{A}}[(\IsSuc[\Var{n}])]}{
\Return{\Ihead & \Var{vs}}{\LiftFold[\IsSucAlg\: \IheadAlg] \quad\where}
} \\
\begin{array}{@{\:\:}l}
   \Let{\IheadAlg & \PiTel{\Var{vs}}{\InterpretIDesc{\IDescOf{\Vector{}}}[(\Lam{\Var{n'}}\IMaybe{\Var{A}}[(\IsSuc[\Var{n'}])])]\: \Var{n}}}
       {}{
   \multicolumn{4}{l}{\qquad\qquad\IMaybe{\Var{A}}[(\IsSuc[\Var{n}])]}\\
   \Return{\IheadAlg & \Tag{\CN{nil}}}{\Hole}
   \Return{\IheadAlg & (\Tag{\CN{cons}}\: \Var{a}\: \Var{xs})}{\Hole}
\end{array}
}}
\]
}
Once again, it is beyond the scope of this paper to formalise the
elaboration process from the high-level notation to the low-level type
theory. The reader will convince himself that the high-level notation
contains all the information necessary to conduct this task. We shall
now freely use the high-level syntax, with the understanding that it
automatically builds a type theoretic term that type-checks.

\end{example}

\begin{example}[Transporting the recursion pattern of \(\_\Le\_\)]

To implement \(\Ilookup\), we use \(\LiftCase\) to transport the case
analysis on \(n\) and \(\LiftInd\) to transport the induction over
\(m\). In a high-level notation, this interaction results in:
{
\[
\Case{
\multicolumn{5}{@{}l}{\Ilookup \::\: \Patch[\LeType\: \LookupType\: \_\Le\_]} \\
\LiftBy{\Ilookup & 
           \Var{m} & 
           \Var{m^m} & 
           \Var{n} & 
           \Var{vs}}
        {\LiftCase}{
\HoleCommand{\quad \Ilookup & 
            \Var{m} & 
            \Var{m^m} & 
            \Zero & 
            \VNil}
\LiftBy{\quad \Ilookup & 
             \Var{m} & 
             \Var{m^m} & 
             (\Suc[\Var{n}]) & 
             (\VCons[\Var{a}\: \Var{vs}])}
        {\LiftInd}{
\HoleCommand{\quad\quad \Ilookup & 
            \Zero & 
            \Zero & 
            \Zero & 
            \VNil}
\HoleCommand{\quad\quad \Ilookup & 
            (\Suc[\Var{m}]) & 
            (\Suc[\Var{m^m}]) & 
            \Zero & 
            \VNil}}}}
\]
}
It is crucial to understand here that, in an interactive setting, the
user would type in the \(\DoLiftBy\) command together with the action
to be carried out and the computer would automatically generate the
resulting patterns.

\end{example}

\begin{example}[Transporting the recursion pattern of \(\_\Plus\_\)]

In order to implement \(\VectorAppend\), we can also benefit from our
generic machinery. We simply have to instruct the machine that we want
to lift the case analysis used in the definition of \(\_\Plus\_\) and
we are left filling the following goals:
{
\[
\Case{
\multicolumn{5}{@{}l}{\VectorAppend \::\: \Patch[\PlusType\: \AppendType\: \_\Plus\_]} \\
\LiftBy{\VectorAppend & 
           \Var{m} & 
           \Var{xs} & 
           \Var{n} & 
           \Var{ys}}
        {\LiftCase}{
\HoleCommand{\quad \VectorAppend & 
            \Zero & 
            \VNil & 
            \Var{n} & 
            \Var{ys}}
\HoleCommand{\quad \VectorAppend & 
            (\Suc[\Var{m}]) & 
            (\VCons[\Var{a}\: \Var{xs}]) & 
            \Var{n} & 
            \Var{ys}}}}
\]
}

\end{example}


\subsection{Transporting constructors}


Just as the recursive structure, the returned values often simply
mirror the original definition: we are in a situation where the base
function returns a given constructor and we would like to return its
ornamented counterpart. Informing the computer that we simply want to
lift the constructor, the computer should fill in the parts that are
already determined by the original constructor and ask only for the
missing information, \ie the data freshly introduced by the ornament.

Remember that, when implementing the coherent lifting, we are working
on the reornaments of the lifting type. Hence, when returning a
constructor-headed value, we are building an inhabitant of a
reornament. When defining reornaments in Section~\ref{sec:reornament},
we have shown that, thanks to deletion ornaments, a reornament can be
decomposed in two components:
\begin{itemize}
\item first, the extension that contains all the extra information
  introduced by the ornament ;
\item second, the recursive structure of the refined datatype, which
  defines the type of the arguments of the constructor
\end{itemize}
And no additional information is required: all the information
provided by indexing with the unornamented datatype is optimally used
in the definition of the reornament. There is absolutely no
duplication of information.

This clear separation of concern is a blessing for us: when lifting a
constructor, we only have to provide the extension and the arguments
of the datatype, nothing more. In term of implementation, this is as
simple as:
{
\[
\Code{
\begin{array}{@{}l@{\:}l@{\:\:}l}
\LiftConstructor &
     \PiTel{\Var{xs}}{\InterpretIDesc{\Var{D}\: (\Var{re}\: \Var{j})}[\IMu[\Var{D}]]} \\
   & \PiTel{\Var{e}}{\ReornExtension[(\Var{O}\: \Var{j})\: \Var{xs}]} & \hspace*{-2em}\CommentLine{coherent extension} \\
   & \PiTel{\Var{a}}{\InterpretIDesc{\Interpretorn{\ReornStructure[\Var{O}\: \Var{xs}\: \Var{e}]}}%
                                    [(\IMu[\OrnAlgOrn{\Var{D}}{\Var{O}}])]} & \CommentLine{arguments}\\
   & \PiTel{\Var{t^{++}}}{\Patch[\Var{T}\: \Var{t}\: \Var{T^+}]} \\
   & : \begin{array}[t]{@{}l@{\:}l}
        \Patch & (\muTimes{\Var{D}}{(\Var{re}\: \Var{j})}{\Var{T}}) \\
               & \Pair{\In[\Var{xs}]}{\Var{t}} \\
               & (\muOrnTimes{\Var{O}}{\Var{j}}{\Var{T^+}})
       \end{array} \\
\end{array} \\
\LiftConstructor[\Var{xs}\: \Var{e}\: \Var{a}\: \Var{t^{++}}] \DoReturn \Pair{\In[\Pair{\Var{e}}{\Var{a}}]}{\Var{t^{++}}}
}
\]
}



\newcommand{\LiftUnit}{\green{\blacksquare}}

\begin{example}[Transporting the constructors of \(\IsSuc\)]

Let us finish the implementation of \(\Head\) from \(\IsSuc\). Our
task is simply to transport the \(\True\) and \(\False\) constructors
along the \(\Maybe{}\) ornament. In a high-level notation, we would
write the command shown in Fig.~\ref{fig:ihead-implem}(c). The
interactive system would then respond by generating the code of
Fig.~\ref{fig:ihead-implem}(d). The \(\Unit\) goals are trivially solved,
probably automatically by the system. The only information the user
has to provide is a value of type \(A\) returned by the \(\Just\)
constructor.

\end{example}

\begin{example}[Transporting the constructors of \(\_\Le\_\)]


In the implementation of \(\Ilookup\), we want to lift the returned
\(\True\) and \(\False\) to the \(\Maybe{}\) ornament. In a high-level
notation, this would be represented as follows:
{
\[
\Case{
\Let{\Ilookup & & & &}{\Patch[\LeType\: \LookupType\: \_\Le\_]}{
\LiftBy{\Ilookup & 
           \Var{m} & 
           \Var{m^m} & 
           \Var{n} & 
           \Var{vs}}
        {\LiftCase}{
\LiftReturn{\quad \Ilookup & 
            \Var{m} & 
            \Var{m^m} & 
            \Zero & 
            \VNil}
           {\Nothing\: \Void}{\Void}{}
\LiftBy{\quad \Ilookup & 
             \Var{m} & 
             \Var{m^m} & 
             (\Suc[\Var{n}]) & 
             (\VCons[\Var{a}\: \Var{vs}])}
        {\LiftInd}{
\LiftReturn{\quad\quad \Ilookup & 
            \Zero & 
            \Zero & 
            (\Suc[\Var{n}]) & 
            (\VCons[\Var{a}\: \Var{vs}])}
           {\Just[\HoleAnn{a}{\Var{A}}]}{\Void}{}
\HoleCommand{\quad\quad \Ilookup & 
            (\Suc[\Var{m}]) & 
            (\Suc[\Var{m^m}]) & 
            (\Suc[\Var{n}]) & 
            (\VCons[\Var{a}\: \Var{vs}])}}}}
}
\]
}
As before, in an interactive setting, the user would instruct the
machine to execute the command \(\DoLiftReturn\) and the computer
would come back with the skeleton of the expected inputs.  Finishing
the implementation of \(\Ilookup\) is now one baby step away, which we
should jump straightaway:
{
\[
\Let{\Ilookup & & & &}{\Patch[\LeType\: \LookupType\: \_\Le\_]}{
\LiftBy{\Ilookup & 
           \Var{m} & 
           \Var{m^m} & 
           \Var{n} & 
           \Var{vs}}
        {\LiftCase}{
\LiftReturn{\quad \Ilookup & 
            \Var{m} & 
            \Var{m^m} & 
            \Zero & 
            \VNil}
           {\Nothing\: \Void}{\Void}{}
\LiftBy{\quad \Ilookup & 
             \Var{m} & 
             \Var{m^m} & 
             (\Suc[\Var{n}]) & 
             (\VCons[\Var{a}\: \Var{vs}])}
        {\LiftInd}{
\LiftReturn{\quad\quad \Ilookup & 
            \Zero & 
            \Zero & 
            (\Suc[\Var{n}]) & 
            (\VCons[\Var{a}\: \Var{vs}])}
           {\Just[\Var{a}]}{\Void}{}
\Return{\quad\quad \Ilookup & 
            (\Suc[\Var{m}]) & 
            (\Suc[\Var{m^m}]) & 
            (\Suc[\Var{n}]) & 
            (\VCons[\Var{a}\: \Var{vs}])}
       {\Ilookup[\Var{m}\: \Var{m^m}\: \Var{n}\: \Var{vs}]}}}}
\]
}

\end{example}

\begin{example}[[Transporting the constructors of \(\_\Plus\_\)]

We can also benefit from the automatic lifting of constructors to fill
out the \(\VCons\) case of vector append. We instruct the
system that we want to lift the \(\Suc\) constructor and get the
following goals as a result:
{
\[
\Case{
\multicolumn{5}{@{}l}{\VectorAppend \::\: \Patch[\PlusType\: \AppendType\: \_\Plus\_]} \\
\LiftBy{\VectorAppend & 
           \Var{m} & 
           \Var{xs} & 
           \Var{n} & 
           \Var{ys}}
        {\LiftCase}{
\HoleCommand{\: \VectorAppend & 
            \Zero & 
            \VNil & 
            \Var{n} & 
            \Var{ys}}
\LiftReturn{\: \VectorAppend & 
            (\Suc[\Var{m}]) & 
            (\VCons[\Var{a}\: \Var{xs}]) & 
            \Var{n} & 
            \Var{ys}}}
           {\VCons\: \HoleAnn{\Var{A}}}
           {\Hole}
           {}
}
\]
}
It is then straightforward to, manually this time, conclude the
implementation of \(\VectorAppend\):
{
\[
\Case{
\multicolumn{5}{@{}l}{\VectorAppend \::\: \Patch[\PlusType\: \AppendType\: \_\Plus\_]} \\
\LiftBy{\VectorAppend & 
           \Var{m} & 
           \Var{xs} & 
           \Var{n} & 
           \Var{ys}}
        {\LiftCase}{
\Return{\: \VectorAppend & 
            \Zero & 
            \VNil & 
            \Var{n} & 
            \Var{ys}}{\Var{ys}}
\LiftReturn{\: \VectorAppend & 
            (\Suc[\Var{m}]) & 
            (\VCons[\Var{a}\: \Var{xs}]) & 
            \Var{n} & 
            \Var{ys}}}
           {\VCons\: \Var{a}}
           {\VectorAppend[\Var{m}\: \Var{xs}\: \Var{n}\: \Var{ys}]}
           {}
}
\]
}

\end{example}

\section{Removing index-level computations}
\label{sec:remove-index-computation}


For a seasoned programmer, the type of \(\Ilookup\) might appear
rather unconventional:
\[
\TypeAnn{\Ilookup}{\PiTel{\Var{m}}{\Nat}
                   \PiTo{\Var{vs}}{\Vector{\Var{A}}[\Var{n}]} 
                     \IMaybe{\Var{A}}[(\Var{m} \Le \Var{n})]}
\]
Indeed, lookup in a vector is traditionally given the following type:
\[
\TypeAnn{\Vlookup}
        {\PiTel{\Var{m}}{\Fin[\Var{n}]}
         \PiTo{\Var{vs}}{\Vector{\Var{A}}[\Var{n}]}
         \Id{\Var{A}}}
\]

The issue with the first presentation is that the return type is
indexed by a \emph{computed} value, \(m \Le n\). This means that,
provided that we have an \(\TypeAnn{x}{\IMaybe{\Var{A}}[k]}\) and we
want to return it, we must first make sure that the computation of \(m
\Le n\) has unfolded to \(k\), where \(k\) might be a value (such as
\(\True\) or \(\False\)) or a suspended computation (such as \(\Suc[m]
\Le n\)). Making sure that this index-level computation unfolds
correctly can be cumbersome at best, and sometimes simply
impossible. One then has to manually rewrite the goal using
proofs. Again, this pollutes the function definition with
computationally unnecessary details.

On the other hand, the second presentation requires absolutely no
special care to indexes, nor any need for proofs. The key difference
is in handling the constraint that \(m\) must be less than the length
of the vector. In the first case, we make no assumption on \(m\) as an
input and then constrain the result to be meaningful if \(m \Le n\)
and pointless otherwise (by definition of \(\IMaybe{}\)). In the
second case, we restrict the input \(m\) to be less than \(n\) (by
stating that \(\TypeAnn{m}{\Fin[n]}\)) and therefore the result is
always meaningful.


\subsection{From \(\Ilookup\) to \(\Vlookup\), manually}

\Spacedcommand{\Truth}{\Function{1}}
\newcommand{\DTo}{\stackrel{\blue{\cdot}}{\To}}

\trashtalk{That's rather shaky because they are not exactly
  equivalent: the second half has been simplified because it is
  trivial with the \(\False\) index}

It is intuitively clear that these two presentations are
equivalent. However, sometimes it is easier to work with the first
(\eg because we use our lifting machinery that, by design, unfolds
the indexed computation as wanted) or with the second (\eg the
recursion pattern we want does not mirror the one of the indexed
computation). Let us look closer at this equivalence, factoring out
the vector that has no influence here:
\[
\PiTo{\Var{m}}{\Nat} \IMaybe{\Var{A}}[(\Var{m} \Le \Var{n})]
  \cong
\Fin[\Var{n}] \To \Id{\Var{A}}
\]
While this seems to suggest an adjunction, it is not clear what the
functors are. Rewriting the equation more abstractly help to see
clearer:
\[
\Truth[\Nat] \DTo \IMaybe{\Var{A}}[(\_ \Le \Var{n})]
  \cong
\Fin[\Var{n}] \To \Id{\Var{A}}
\]
Where \(\Truth\) is the map from \(\Nat\) to \(\Unit\) and \(P \DTo
Q\) is defined as \(\PiTo{\Var{x}}{\Nat} P\: \Var{x} \To Q\:
\Var{x}\). Having done that, we can make a more informed guess on the
right adjoint: it is the functor \(\LamAnn{\Var{X}}{\Nat \To
  \Set}{\LamAnn{\Var{m}}{\Nat}{\Var{X} \circ (\_ \Le \Var{n})}}\).

Now, as for the left adjoint, we must extract a functor that would map
\(\Truth[\Nat]\) to \(\Fin[n]\). The semantics of \(\Fin[n]\) is of
great help here: intuitively, an inhabitant of \(\Fin[n]\) is a number
\(\TypeAnn{m}{\Nat}\) such that \(m \Le n\), or put otherwise
\[
\Fin[n] \cong \{ m \in \Nat \:|\: m \Le n\}
\]
As we have seen in Section~\ref{sec:algebraic-ornament}, these two
presentations are equivalent: seen as the algebraic ornament by the
algebra of \(\_ \Le n\), \(\Fin[n]\) can be decomposed as a number (as
computed by \(\OrnForget{\OrnAlgebraic{\Nat}{\alpha_{\Le}}}\)) and a
proof (as computed by \(\OAAO\)). Following
\citet{atkey:inductive-refinement}, we translate this into category,
obtaining that \(\Fin[n] \cong \Sigma_{(\_ \Le n)} \circ
\Truth[\Nat]\). This suggests the following equivalences:
\begin{align*}
\Truth[\Nat] \DTo \IMaybe{\Var{A}}[(\_ \Le \Var{n})]
  &\cong
\Fin[\Var{n}] \To \Id{\Var{A}} \\
  &\cong
\Sigma_{(\_ \Le n)} \Truth[\Nat] \DTo \IMaybe{\Var{A}}
\end{align*}
The second and third line are indeed equivalent: the third line maps
predicates over \(\Bool\), which might seem as more general than the
second line, however in the \(\False\), \(\IMaybe{}\) simplifies to
\(\Unit\). Hence, the only interesting case happens when the index is
\(\True\), in which case \(\IMaybe{}\) simplifies to
\(\Id{}\). \trashtalk{That's a lie here because \(\_ \Le n\) is
  not defined by a fold in the implementation I gave. So there is a
  lot of machinery that won't work.}

With this more abstract presentation, it should be clear what the left
adjoint must be: it is \(\Sigma_{(\_ \Le n)}\). Now we have to check
that \(\Sigma_{(\_ \Le n)}\) is indeed left adjoint to \(\_ \circ (\_
\Le \Var{n})\): it is actually true in any locally cartesian
closed-category (LCCC). Hence, a fortiori, this is true in type
theory, as type-theories are a model of LCCCs. \trashtalk{Careful
  here, risk of injury.}

\subsection{Reindexing and algebraic ornaments}

In all generality, we prove the following equivalence:
\[
\PiTel{\Var{m}}{\IMu[D]} \To \IMu[E\: (\Fold{\alpha}\: m)]
  \cong
\IMu[\OrnAlgebraic{D}{\alpha}\: n] \To \IMu[E\: n]
\]
The proof is straightforward, following the intuition we gave above:
\begin{align*}
\PiTel{\Var{m}}{\IMu[D]} \To \IMu[E\: (\Fold{\alpha}\: m)]
  & \cong \Truth{\IMu[D]} \DTo \IMu[E] \circ \Fold{\alpha} & \mbox{(abstract the index)} \\
  & \cong \Sigma_{\Fold{\alpha}} \Truth{\IMu[D]} \DTo \IMu[E] & (\Sigma_f \dashv \_ \circ f) \\
  & \cong \IMu[\Interpretorn{\OrnAlgebraic{D}{\alpha}}] \DTo \IMu[E] & \mbox{(coherence of algebraic ornament)}
\end{align*}

\trashtalk{That's not very serious: I give a lengthy example, then to
  prove a triviality.}


\Spacedcommand{\RightToLeftAdjoint}{\Function{rlAdjoint}}
\Spacedcommand{\LeftToRightAdjoint}{\Function{lrAdjoint}}

Through our constructive glasses, this categorical equivalence
correspond to two rather interesting functions: in one direction,
\(\RightToLeftAdjoint\) transforms a function with index-level
computation into one with a stronger premise ; in the other direction,
\(\LeftToRightAdjoint\) turns a function with strong premises into one
with index-level computation. The implementation is absolutely
unsurprising: we translate the categorical proof into its
corresponding constructive operation, \ie the \(\OAAO\) theorem and
the \(\OrnMake{}{}\) function:
\[
\Code{
\begin{array}{@{}l@{\:}l}
\RightToLeftAdjoint & 
     \PiTel{\Var{f}}
           {\PiTel{\Var{i}}{\Var{I}} 
                \PiTo{\Var{t}}{\IMu[\Var{D}\: \Var{i}]}
                \IMu[\Var{E}\: \Pair{\Var{i}}{\Fold{\Var{\alpha}}\: \Var{t}}]} \\
   & \PiTel{\Var{i}}{\Var{I}} 
     \PiTel{\Var{x}}{\Var{X}\: \Var{i}} 
     \PiTel{\Var{t^x}}{\IMu[\Interpretorn{\OrnAlgebraic{\Var{D}}{\Var{\alpha}}}\: \Pair{\Var{i}}{\Var{x}}]} 
   : \IMu[\Var{E}\: \Pair{\Var{i}}{\Var{x}}] \\
\RightToLeftAdjoint & \Var{f}\: \Var{i}\: \Var{x}\: \Var{t^x} \DoReturn
       \Subst[(\Lam{\Var{x}}{\IMu[\Var{E}\: \Pair{\Var{i}}{\Var{x}}]})\:
               (\Symmetry[(\OAAO[\Var{i}\: \Var{x}\: \Var{t^x}])])\:
               (\Var{f}\: (\OrnForget{\OrnAlgebraic{\Var{D}}{\Var{\alpha}}}\: \Var{t^x}))]
\end{array}
\\
\\
\begin{array}{@{}l@{\:}l}
\LeftToRightAdjoint & 
     \PiTel{\Var{g}}
           {\PiTel{\Var{i}}{\Var{I}} 
            \PiTel{\Var{x}}{\Var{X}\: \Var{i}} 
            \PiTo{\Var{t^x}}{\IMu[\Interpretorn{\OrnAlgebraic{\Var{D}}{\Var{\alpha}}}\: \Pair{\Var{i}}{\Var{x}}]}
            \IMu[\Var{E}\: \Pair{\Var{i}}{\Var{x}}]} \\
   & \PiTel{\Var{i}}{\Var{I}} 
     \PiTel{\Var{t}}{\IMu[\Var{D}\: \Var{i}]}
   : \IMu[\Var{E}\: \Pair{\Var{i}}{\Fold{\Var{\alpha}}\: \Var{t}}] \\
\LeftToRightAdjoint & \Var{g}\: \Var{i}\: \Var{t}\: \DoReturn \Var{g}\: (\OrnMake{\Var{D}}{\Var{\alpha}}\: \Var{t})
\end{array}}
\]

We should finally add that this construction is in no sense restricted
to the functions computed by \(\Patch\): the relation we have
established and made concrete here can be used for any function. Of
general interest is the function \(\LeftToRightAdjoint\): instead of
tediously unfolding the computation \(\Fold{\alpha}\), the programmer
can instead build the alternative function without having to care
about the recursion pattern introduced by \(\alpha\).

In our framework, the \(\RightToLeftAdjoint\) is more interesting: our
clever constructors make it easy to unfold the index-level
computation, however, in practice, one will rather \emph{use} the
equivalent function using richer data-types. Hence, our user can
implement the \(\Patch\) using our machinery while still getting the
more convenient function.

\trashtalk{But this is saying nothing about the case where the
  index-level function takes many index-arguments} \trashtalk{Also, it
  seems that there is a story of weakest precondition/strongest
  postcondition here, with both being adjoint/there being a Galois
  connection between the two notions.}


\paragraph{Example: from \(\Ilookup\) to \(\Vlookup\).}


Using the gadget we have just developed, we can now obtain
\(\Vlookup\) automatically from the definition of \(\Ilookup\):
\[
\Let{\Vlookup & \PiTel{\Var{m}}{\Fin[\Var{n}]} & \PiTel{\Var{vs}}{\Vector{\Var{A}}[\Var{n}]}}{\Id{\Var{A}}}{
\Return{\Vlookup & \Var{m} & \Var{vs}}
       {\RightToLeftAdjoint[(\Lam{\Var{m}}{\Ilookup\: \Var{m}\: \Var{vs}}) 
                            \Void\: 
                            \True\: 
                            m]}
}
\]


\section{Related work}


Our work is an extension of the work of \citet{mcbride:ornament} on
ornaments, originally
introduced to organise datatypes according to their common
structure. This gave rise to the notion of ornamental algebras --
forgetting the extra information of an ornamented datatype -- and
algebraic ornaments -- indexing a datatype according to an algebra.
This, in turn, induced the notion of algebraic ornament by
ornamental algebras, which is a key ingredient for our work.

However, for simplicity of exposition, these ornaments had originally
been defined on a less index-aware universe of datatypes. As a
consequence, computation over indices was impossible and, therefore,
deletion of duplicated information was impossible. A corollary of this
was that reornaments contained a lot of duplication, hence making the
lifting of value from ornamented to reornamented datatype extremely
tedious.


Our presentation of algebraic ornament has been greatly improved by
the categorical model developed by \citet{atkey:inductive-refinement}:
the authors gave a conceptually clear treatment of algebraic ornament
in a Lawvere fibration. At the technical level, the authors connected
the definition of algebraic ornament with truth-preserving liftings,
which are also used in the construction of induction principles, and
op-reindexing, which models \(\Sigma\)-types in type theory.

Whilst the authors did not explicitly address the issue of
transporting functions across ornaments, much of the infrastructure
was implicitly there: for instance, lifting of folds is a trivial
specialisation of induction. Also, the characterisation of the
fix-point of an algebraic ornament as op-reindexing of the fold is a
key ingredient to understanding index-level computations and
assimilate them at the term level.


In their work on realisability and parametricity for Pure Type
Systems, Bernardy and Lasson~\citep{bernardy:realizability} have shown
how to build a logic from a programming language. In such a system,
terms of type theory can be precisely segregated based on their
computational contribution and their logical contribution. In
particular, the idea that natural numbers realise lists of the
corresponding length appears in this system under the guise of
vectors, the reflection of the realisability predicate. The strength
of the realisability interpretation is that it is naturally defined on
functions: while \citet{mcbride:ornament} and
\citet{atkey:inductive-refinement} only consider ornaments on datatypes,
their work is the first, to our knowledge, to capture a general notion
of functions realising -- \ie ornamenting -- other functions.


Following the steps of Bernardy, \citet{ko:modularising-inductive}
adapted the realisability interpretation to McBride's universe of
datatypes and explored the other direction of the \(\Patch\)
equivalence, using reornaments to generate coherence properties: they
describe how one could take list append together with a proof that it
is coherent with respect to addition and obtain the vector append
function. Their approach would shift neatly to our index-aware setting, where
the treatment of reornaments is streamlined by the availability of deletion.

However, we prefer to exploit the direction of the equivalence which
internalises coherence: we would rather use the full power of
dependent types to avoid explicit proof. Hence, in our framework, we
simultaneously induce list append and implicitly prove its coherence
with addition just by defining vector append. Of course, which approach is
appropriate depends on one's starting point.

Moreover, our universe of functions takes a step beyond the related
work by supporting the mechanised construction of liftings, leaving to
the user the task of supplying a minimal patch. Our framework could
easily be used to mechanise the realisability predicate constructions of
\citet{bernardy:realizability}, \citet{ko:modularising-inductive}.


\section{Conclusion}


In this paper, we have developed the notion of functional ornament
and shown how one can achieve code reuse by transporting functions
along a functional ornament. To this end, we have adapted McBride's
ornaments to our universe of datatypes~\citep{dagand:levitation}. This
gave us the ability to compute over indices, hence introducing the
deletion ornament. Deletion ornaments are a key ingredient to the
internalisation of Brady's optimisation~\citep{brady:optimisations}
over inductive families. In particular, this gave us a simpler
implementation of reornaments.


We then generalised ornaments to functions: from a universe of
function type, we define a functional ornament as the ornamentation of
each of its inductive components. A function of the resulting type will
be subject to a coherence property, akin to the ornamental forgetful
map of ornaments. We have constructively presented this object, by
building a small universe of functional ornaments. 


Having functional ornaments, this raises the question of transporting
a function to its ornamented version in such way that the coherence
property holds. Instead of asking our user to write cumbersome proofs,
we defined a \(\Patch\) type as the type of all the functions that
satisfies the coherence property by construction. Hence, we make
extensive use of the dependently typed programming machinery offered by
the environment: in this setting, the type-checker, that is the
computer, is working with us to construct a term, not waiting for us
to produce a proof.


Having implemented a function correct by construction, one then get,
for free, the lifting and its coherence certificate. This is a
straightforward application of the equivalence between \(\Patch\) type
and the set of coherent functions. These projection functions have
been implemented in type theory by simple generic programming over the
universe of functional ornaments.


To further improve code reuse, we provide two clever constructors to
implement a \(\Patch\) type: the idea is to use the structure of the
base function to guide the implementation of the coherent
lifting. Hence, if the base function uses a specific induction
principle or returns a specific constructor, we make it possible for
the user to specify that she wants to lift this element one level
up. This way, the function is not duplicated: only the new
information, as determined by the ornament, is necessary.


To conclude, we believe that this is a first yet interesting step
toward code reuse for dependently typed programming systems. With
ornaments, we were able to organise datatypes by their
structure. With functional ornaments, we are now able to organise
functions by their structure-preserving computational
behaviour. Besides, we have developed some appealing automation to
assist the implementation of functional ornaments, without any proving
required, hence making this approach even more accessible.



\subsection{Future work}



Whilst we have deliberately chosen a simple universe of functions, we
plan to extend it in various directions. Adding type dependency
(\(\Pi\)- and \(\Sigma\)-types) but also non inductive sets is a
necessary first step. Inspired by \citet{bernardy:thesis} but also
\citet{miquel:icc}, we would like to add a parametric quantifier: in
the implementation of \(\Ilookup\), we would like to mark the index
\(A\) of \(\Vector{A}\) and \(\IMaybe{A}\) as parametric so that in
the \(\VCons[a]\) case, the \(a\) could automatically be carried over to
\(\Just[a]\).

The universe of functional ornaments could be extended as well,
especially once the universe of functions has been extended with
dependent quantifiers. For instance, we want to consider the
introduction and deletion of quantifiers, as we are currently doing on
datatypes. Whilst we have only looked at least fixed points in this
paper, we also want to generalise our universe with greatest fixed
points and the lifting of co-inductive definitions.

Further, our framework relies crucially on the duality between a
reornament and its ornament presentation subject to a proof. We cross
this isomorphism in both directions when we project the lifting from
the coherent lifting. In practice, this involves a traversal of each
of the input datatypes and a traversal of each of the output
datatypes. However, computationally, these traversal are identities:
the only purpose of these terms is at the logical level, for the
type-checker to fix the types. We are looking at transforming our
library of clever constructor into a proper domain-specific language
(DSL). This way, implementing a coherent lifting would consists in
working in a DSL. Projecting the lifting and its coherence proof would
then be the work of an optimising compiler that would compute away
the useless translations.

Finally, much work remains to be done on the front of usability: for
convenience, we have presented some informal notations for datatypes,
their ornaments and an extension of Epigram programming facility with
liftings. A formal treatment of these syntaxes and of their
elaboration to the low-level type theory is underway.

\paragraph{Acknowledgements} 
We owe many thanks to our colleagues in the MSP group for many
fruitful discussions. We are in particular grateful to Guillaume
Allais, Stevan Andjelkovic and Peter Hancock for their meticulous
reviews of several drafts of this paper. We shall also thank Edwin
Brady for suggesting the family of lookup functions as a prolific
source of functional ornaments and Andrea Vezzosi for spotting an
issue in our definition of reornaments. Finally, this paper would
have remained a draft without the help and encouragement of Jos\'{e} Pedro
Magalh\~{a}es.


\bibliographystyle{abbrvnat}
\bibliography{paper}



\end{document}